\UseRawInputEncoding
\documentclass[aps,prb,twocolumn,superscriptaddress,showpacs]{revtex4-1}
\usepackage{graphicx}% Include figure files
\usepackage{dcolumn}% Align table columns on decimal point
\usepackage{bm}% bold math
\usepackage{xcolor}
\newcommand{\tc}{$T_c$}
\newcommand{\ef}{$E_F$}
\newcommand{\kf}{$k_F$}
\newcommand{\kz}{$k_z$}

\newcommand{\dxy}{$d_{xy}$}
\newcommand{\dxz}{$d_{xz}$}
\newcommand{\dyz}{$d_{yz}$}
\newcommand{\dxxyy}{$d_{x^2-y^2}$}

\newcommand{\g}{$\Gamma$}
\newcommand{\sto}{SrTiO$_3$}
\newcommand{\gx}{$\Gamma$-X}
\begin{document}

%\preprint{APS/123-QED}

\title{Flatband-Induced Itinerant Ferromagnetism in RbCo$_2$Se$_2$}% Force line breaks with \\

\author{Jianwei Huang}
\affiliation{Department of Physics and Astronomy, Rice University, Houston, TX 77005, USA}
\author{Zhicai Wang}
\affiliation{School of the Gifted Young, University of Science and Technology of China, Hefei, Anhui 230026, China}
\author{Hongsheng Pang}
\affiliation{School of the Gifted Young, University of Science and Technology of China, Hefei, Anhui 230026, China}
\author{Han Wu}
\affiliation{Department of Physics and Astronomy, Rice University, Houston, TX 77005, USA}
\author{Huibo Cao}
\affiliation{Neutron Scattering Division, Oak Ridge National Laboratory, Oak Ridge, Tennessee 37831, USA}
\author{Sung-Kwan Mo}
\affiliation{Advanced Light Source, Lawrence Berkeley National Lab, Berkeley, CA 94720, USA}
\author{Avinash Rustagi}
\affiliation{Department of Physics, North Carolina State University, Raleigh, NC 27695, USA}
\affiliation{School of Electrical and Computer Engineering, Purdue University, West Lafayette, IN 47907}
\author{A. F. Kemper}
\affiliation{Department of Physics, North Carolina State University, Raleigh, NC 27695, USA}
\author{Meng Wang}
\email{wangmeng5@mail.sysu.edu.cn}
\affiliation{School of Physics, Sun Yat-Sen University, Guangzhou, Guangdong 510275, China}
\affiliation{Department of Physics, University of California, Berkeley, CA 94720, USA}
\author{Ming Yi}
\email{mingyi@rice.edu}
\affiliation{Department of Physics and Astronomy, Rice University, Houston, TX 77005, USA}
\affiliation{Department of Physics, University of California, Berkeley, CA 94720, USA}
\author{R. J. Birgeneau}
\email{robertjb@berkeley.edu}
\affiliation{Department of Physics, University of California, Berkeley, CA 94720, USA}
\affiliation{Materials Science Division, Lawrence Berkeley National Laboratory, Berkeley, California 94720,  USA}

%\author{Jianwei Huang$^{1}$, Zhicai Wang$^{2}$, Hongsheng Pang$^{2}$, Han Wu$^{1}$ Sung-Kwan Mo$^{3}$, Avinash Rustagi$^{4}$, A. F. Kemper$^{4}$, R. J. Birgeneau$^{5*}$, Meng Wang$^{6,5*}$, and Ming Yi$^{1,5*}$}
% \affiliation{
%\\$^{1}$Department of Physics and Astronomy, Rice University, Houston, TX 77005, USA
%\\$^{2}$School of the Gifted Young, University of Science and Technology of %China, Hefei, Anhui 230026, China.
%\\$^{3}$Advanced Light Source, Lawrence Berkeley National Lab, Berkeley, CA %94720, USA
%\\$^{4}$Department of Physics, North Carolina State University, Raleigh, NC %27695, USA
%\\$^{5}$Department of Physics, University of California, Berkeley, CA %94720, USA
%\\$^{6}$School of Physics, Sun Yat-Sen University, Guangzhou, Guangdong %510275, China
%\\$^{*}$To whom correspondence should be addressed: robertjb@berkeley.edu, %wangm@berkeley.edu and mingyi@rice.edu.}


\date{\today}% It is always \today, today,

\begin{abstract}
$A$Co$_2$Se$_2$ ($A$=K,Rb,Cs) is a homologue of the iron-based superconductor, $A$Fe$_2$Se$_2$. From a comprehensive study of RbCo$_2$Se$_2$ via measurements of magnetization, transport, neutron diffraction, angle-resolved photoemission spectroscopy, and first-principle calculations, we identify a ferromagnetic order accompanied by an orbital-dependent spin-splitting of the electronic dispersions. Furthermore, we identify the ordered moment to be dominated by a \dxxyy~flatband near the Fermi level, which exhibits the largest spin splitting across the ferromagnetic transition, suggesting an itinerant origin of the ferromagnetism. In the broader context of the iron-based superconductors, we find this \dxxyy~flatband to be a common feature in the band structures of both iron-chalcogenides and iron-pnictides, accessible via heavy electron doping.
%, leading to ferromagnetic fluctuations in proximity to high temperature superconductivity in the iron-based superconductors.

\end{abstract}

\maketitle

%\tableofcontents

\section{INTRODUCTION}
%Research on ternary compounds crystallize in ThCr$_2$Si$_2$-type structure are of great significant in condensed matter physics due to the interesting physics found in these materials such as magnetisc\cite{Kovnir2010, Shatruk2019}, heavy fermion behavior\cite{Steglich1979, Palstra1985} and high temperature superconductivity\cite{Rotter2008}. Especially in the past twelve yeas, the discovery of high temperature superconductivity in the iron pnictides and iron chalcogenides has led a hot research in this type of materials\cite{Rotter2008, Hsu2008, Guo2010, Wang2012b}. 

The parent compounds of most iron-based superconductors (FeSCs) are collinear antiferromagnetic (AFM) metals~\cite{DeLaCruz2008}, with nearly compensated hole and electron Fermi pockets separated by the AFM wavevector~\cite{Yi2017}. Superconductivity emerges with the suppression of the AFM order~\cite{Kordyuk2012}. In the intercalated iron chalcogenides, A$_x$Fe$_{2-\delta}$Se$_2$ (A=K,Rb,Cs), parent compounds typically exhibit insulating behaviors with a variety of AFM orders~\cite{Dai2015, Wang2015a}. By tuning the iron content, one can achieve iron vacancy-free superconducting phases in A$_x$Fe$_2$Se$_2$ (A=K,Rb,Cs)~\cite{Wang2016}, which exhibit large electron Fermi pockets near the Brillouin zone (BZ) corners.
%, similar to that of monolayer FeSe films grown on \sto~substrates with a superconducting gap opening temperature near 65K~\cite{Liu2012, Tan2013, Lee2014}.

ACo$_2$Se$_2$ (A=K,Rb,Cs) is an isostructural homologue of the vacancy-free superconducting AFe$_2$Se$_2$ phase with Fe substituted by Co~\cite{HUAN1989}, albeit with distinct physical properties. While AFe$_2$Se$_2$ is a superconductor with an AFM insulating parent phase~\cite{Dai2015, Wang2015a, Wang2016}, ACo$_2$Se$_2$ exhibits metallic magnetic ground states without superconductivity. In particular, ACo$_2$Se$_2$ consists of planar ferromagnetic (FM) sheets that are either aligned (KCo$_2$Se$_2$ and RbCo$_2$Se$_2$) or anti-aligned  (A-type AFM in CsCo$_2$Se$_2$)~\cite{Metals1989, HUAN1989, Yang2013, VonRohr2016}. 
%A metamagnetism-like behavior has been reported in CsCo$_2$Se$_2$ in the presence of an external magnetic field~\cite{Yang2013}. 
Due to the metallicity, an itinerant nature has been proposed as the origin of the magnetism~\cite{EdmundCliftonStoner1938, Metals1989}. In such a scenario, band splitting into the spin-majority and spin-minority bands is expected~\cite{Yang2004, Guttler2016, Mazzola2018}. 
However, while the low temperature electronic structure of KCo$_2$Se$_2$ has been measured~\cite{Liu2015b}, no direct observation of electronic reconstruction across the FM transition has been reported for this series of itinerant magnets.
%Previous \yi{angle-resolved photoemission spectroscopy} (ARPES) studies on KCo$_2$Se$_2$ revealed the Fermi surface and band structure in the ferromagnetic state\cite{Liu2015b}. \yi{[since the intro only discusses chalcogenides, i would leave this pnictide example to the discussion.] A flatband around Fermi level was also reported in SrCo$_2$As$_2$ which is in support of the itinerant picture\cite{Li2019}.} However, temperature dependent experimental data are lacked that it is not clear whether and which bands are splitted in the ferromagnetic state. 

Here we report the evolution of the electronic structure of RbCo$_2$Se$_2$ across the FM transition via angle-resolved photoemission spectroscopy (ARPES), together with characterization by magnetization, transport and neutron diffraction measurements.
%The Fermi surfaces (FSs) of RbCo$_2$Se$_2$ in the paramagnetic state consist of one electron pocket around the BZ center, $\Gamma$, and three electron pockets around the BZ corner, X. A comparison with first-principle calculations suggests moderate electron-electron correlations. 
We identify a nearly flatband near the Fermi level (\ef) that exhibits the largest splitting in the FM phase. 
%Temperature-dependent measurements reveals none-zero band splitting persisting to above the FM ordering temperature, indicating possible ferromagnetic fluctuations in the paramagnetic state. 
From first-principle calculations, we identify this band to be a \dxxyy~flatband that contributes the most to the density of states at \ef~and therefore drives the itinerant FM in this material.
% as reached by heavily electron-doping the isostructural superconducting phase RbFe$_2$Se$_2$. 
Furthermore, in the larger context of the FeSCs, we find this \dxxyy~flatband to be a common feature in the calculated band structures in both the FeSe-based and FeAs-based systems that is accessible via heavy electron-doping. Combining the phenomenology across the FeSC families, we point out a connection between the emergence of various symmetry-breaking phases to the common features in the low energy Fe 3d bands tunable via carrier doping.

\section{METHODS}
Single crystals of RbCo$_2$Se$_2$ were grown by the self-flux method~\cite{Yang2013}. High purity Rb, Co and Se were used as the starting materials and prepared in the ratio of 1:2:2. Sample magnetization and resistivity were measured with commercial MPMS and PPMS. Neutron diffraction was carried out on the HB3A four-circle diffractometer at the High Flux Isotope Reactor, Oak Ridge National Laboratory with a neutron wavelength of $\lambda=1.553$\AA~selected by a bent perfect Si-220 monochromator~\cite{Chakoumakos2011}. The nuclear and magnetic structures were refined with the FullProf suite~\cite{Rodriguez-Carvajal1993}, resulting in a refined stoichiometry of Rb$_{0.93}$Co$_{1.87}$Se$_2$, and lattice constants $a =$ 3.870 \AA~ and $c =$ 13.876 \AA. ARPES measurements were carried out at beamline 10.0.1 of the Advanced Light Source with a Scienta R4000 hemispherical energy analyzer. The energy resolution was set as 12.5 meV and the angular resolution was set as 0.3$^\circ$. The photon energy of the light was set at 45 eV in an $s$-polarization geometry where the polarization vector was perpendicular to the electron analyzer slit. The samples were cleaved in situ at 30 K and kept in ultra high vacuum with a base pressure better than 3 $\times$ 10$^{-11}$ torr during measurements.

The electronic structure calculations were performed using density functional theory (DFT) via QUANTUM ESPRESSO~\cite{Giannozzi2009, Giannozzi2017} with plane wave scalar relativistic pseudopotentials. The exchange-correlation energy was described by the generalized gradient approximation in the scheme proposed by Perdew-Burke-Ernzerhof~\cite{PotentialforCalculation} with a wavefunction cutoff energy of 40 Ry. The BZ was sampled for integration according to the scheme proposed by Monkhorst-Pack~\cite{Monkhorst1976} with a grid of 10 $\times$ 10 $\times$ 10 k-points. Experimentally determined lattice constants were used with the out-of-plane Se atomic locations determined from relaxing the system. Calculations were performed for both the non-magnetic and the FM case where the total magnetization per unit cell was constrained to match that of the experimental value.
%1.18 Bohr magneton ($\mu_B$) (corresponding to the two Co atoms, each contributing 0.59 $\mu_B$ as determined experimentally).
\begin{figure}
\includegraphics[width=0.48\textwidth]{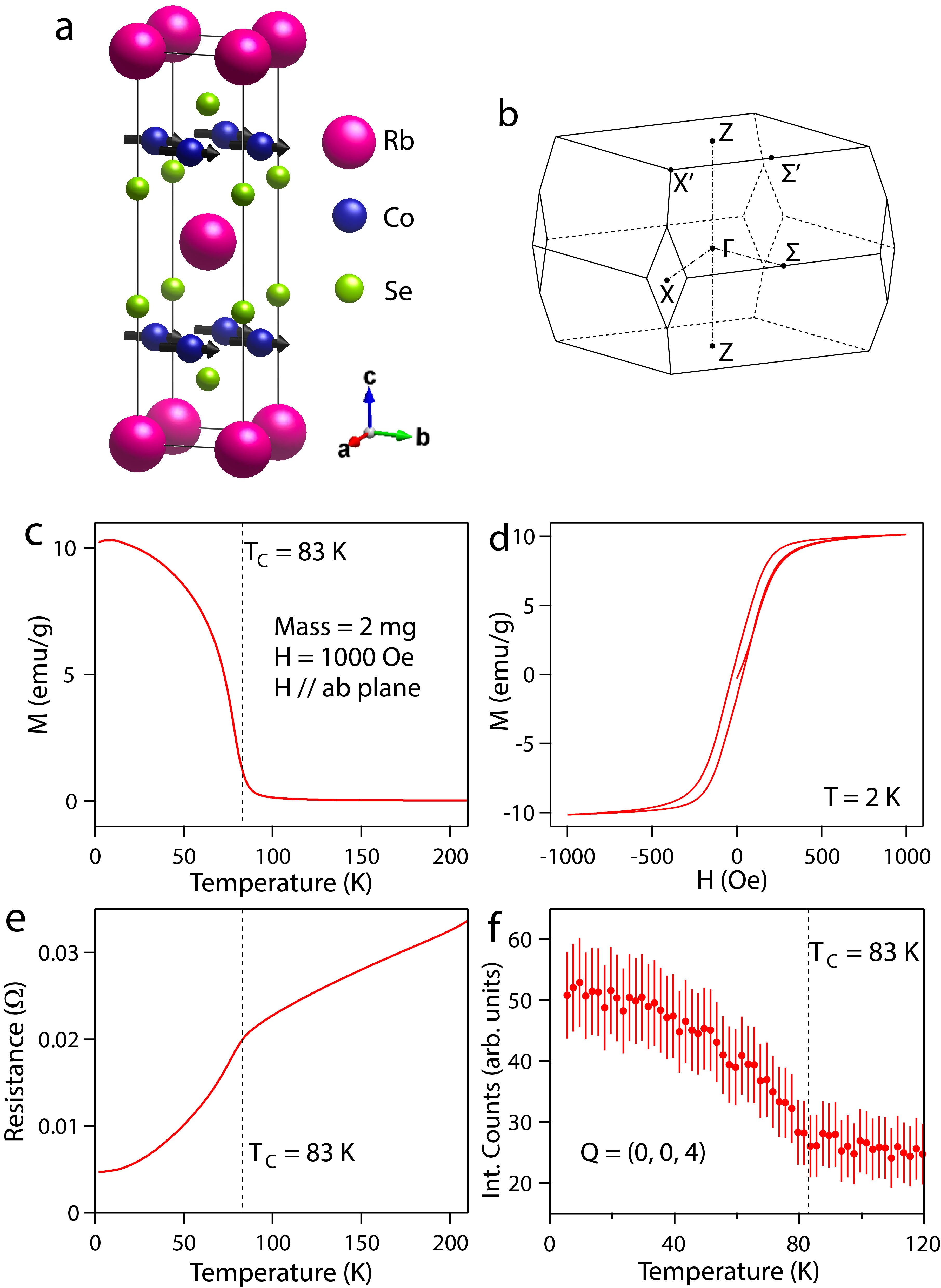}
\caption{\label{fig:Fig1} {\bf FM characterization.} (a) Crystal and magnetic structure of RbCo$_2$Se$_2$ with magnetic moments indicated by arrows. (b) BZ notations. (c) Temperature-dependent magnetization at 1000 Oe external magnetic field. (d) Field-dependent magnetization at 2 K. (e) Resistance as a function of temperature. (f) Neutron diffraction measurements of the integrated intensity of the (0,0,4) magnetic Bragg peak as a function of temperature.}
\end{figure}

\section{RESULTS}

The crystal and magnetic structures of RbCo$_2$Se$_2$ are shown in Fig.~\ref{fig:Fig1}a. The ordered magnetic moment is determined from neutron diffraction to be 0.60(4) $\mu_B$ per Co site aligned along the $a$ axis, consistent with previous reports~\cite{HUAN1989, Yang2013}.
%The corresponding tetragonal BZ notation is shown in Fig.~\ref{fig:Fig1}b. 
Magnetization measurement with the magnetic field in the $ab$ plane identifies an onset of the FM order at \tc~= 83 K (Fig.~\ref{fig:Fig1}c). A hysteresis behavior confirming the FM ground state is observed at 2 K as a function of external field (Fig.~\ref{fig:Fig1}d). A kink at \tc~can also be identified in the resistance measurement (Fig.~\ref{fig:Fig1}e). Finally, the integrated counts at the (0,0,4) magnetic Bragg peak from our neutron diffraction measurements clearly confirm the FM transition at \tc~(Fig. ~\ref{fig:Fig1}f).

\begin{figure}
\centering
\includegraphics[width=0.48\textwidth]{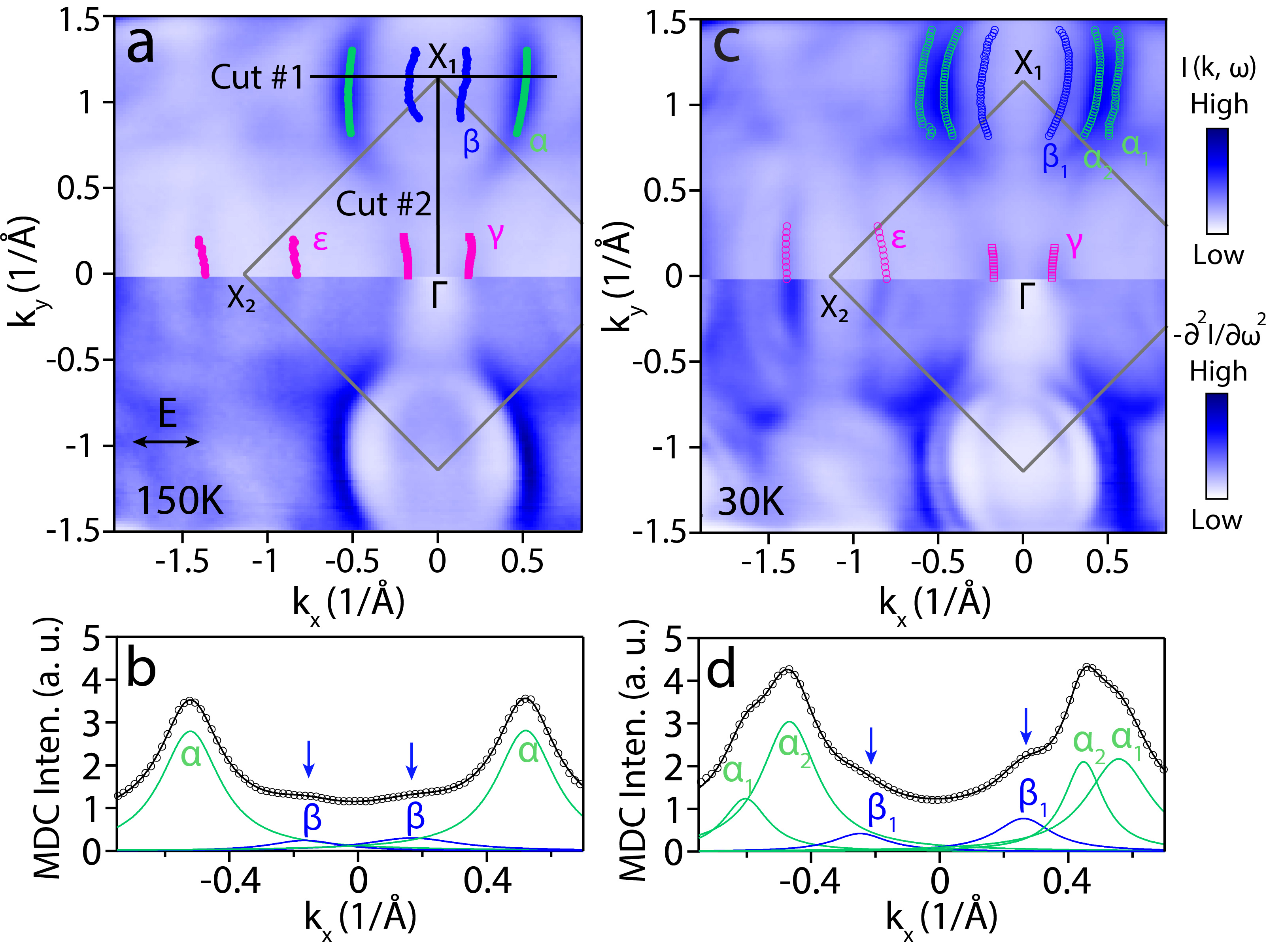}
\caption{\label{fig:Fig2} {\bf Fermi surfaces across \tc.} (a) FS mapping at 150 K $>$ \tc. The $k_y > 0$ ($k_y < 0$) region shows the raw data (2D curvature). Markers indicate MDC-fitted FSs. (b) MDC at \ef~along cut \#1 in (a) fitted with four Lorentzians and a constant background. The blue arrows indicate the $\beta$ band. (c,d) Same as (a,b) except taken at 30 K in the FM state.}
\end{figure}

Having confirmed the FM transition, we present the measured electronic structure in the paramagnetic phase. At 150 K $>$ \tc, the Fermi surfaces (FSs) of RbCo$_2$Se$_2$ exhibit one small electron pocket ($\gamma$) around the BZ center, $\Gamma$, and three electron pockets around the BZ corner, X (Fig.~\ref{fig:Fig2}a). To show better contrast, we present the raw data in the upper half of the FS, and its 2D curvature images in the bottom half. We note that the X$_1$ and X$_2$ points as labeled are equivalent under $C_4$ rotational symmetry of the tetragonal crystal structure. However, the map intensity appears different here due to the usage of linear horizontal polarization, which probes the \dxz~and \dyz~orbitals differently due to photoemission matrix element effects~\cite{Yi2011}. A large electron pocket ($\alpha$) around X$_1$ is easy to discern (Fig.~\ref{fig:Fig2}a). The presence of a second weaker inner electron band ($\beta$) is evident both from the FS images at X$_1$ as well as from the momentum-distribution curve (MDC) taken from \ef, which can be fitted with four Lorentzian peaks on a constant background (Fig.~\ref{fig:Fig2}b). A third electron band ($\varepsilon$) can be observed at the X$_2$ point due to the distinct matrix element effect for the two equivalent X-points (Fig.~\ref{fig:Fig1}a)\cite{Yi2011}. Its fitted \kf~points (Fermi momenta) are plotted in Fig.~\ref{fig:Fig2}a, which are distinct from those of the $\alpha$ and $\beta$ electron bands, confirming a total of three electron pockets at the X point. The identification of their dominant orbital character and expected matrix element effects will be discussed in a later section.
Compared with the iron-chalcogenide superconductors AFe$_2$Se$_2$, the electron pockets of RbCo$_2$Se$_2$ at the X point are much larger, consistent with additional electron doping provided by the substitution of Fe by Co~\cite{Mou2011a, Zhang2011a}.

In comparison, the number of FS sheets observed deep in the FM phase increases. Most notable is the splitting of the $\alpha$ FS at X$_1$ (Fig.~\ref{fig:Fig2}c). This is evident both from the FS map in Fig.~\ref{fig:Fig2}c and a comparison of the MDCs across X$_1$ (Fig.~\ref{fig:Fig2}d), which can now be fitted with six Lorentzian peaks showing a splitting of the peak previously identified as the $\alpha$ pocket at 150 K. This is consistent with a band splitting due to the FM ordering. We therefore label these as the $\alpha_1$ and $\alpha_2$ pockets, which correspond to the spin-majority and spin-minority bands, respectively. The \kf~opening (defined by the separation in momentum) of the pair of peaks labeled $\beta_1$ has expanded compared to that of the $\beta$ pocket in the paramagnetic phase, suggesting that the pair observed at 30 K is likely the spin-majority branch of the $\beta$ band that has shifted downwards in energy while the spin minority $\beta_2$ band has shifted to above \ef~and hence is not observed. In contrast, we do not observe obvious shifts of the electron pocket ($\gamma$) at $\Gamma$ and the electron pocket ($\varepsilon$) at X$_2$ below \tc (Fig.~\ref{fig:Fig2}a, b).

\begin{figure}
\centering
\includegraphics[width=0.48\textwidth]{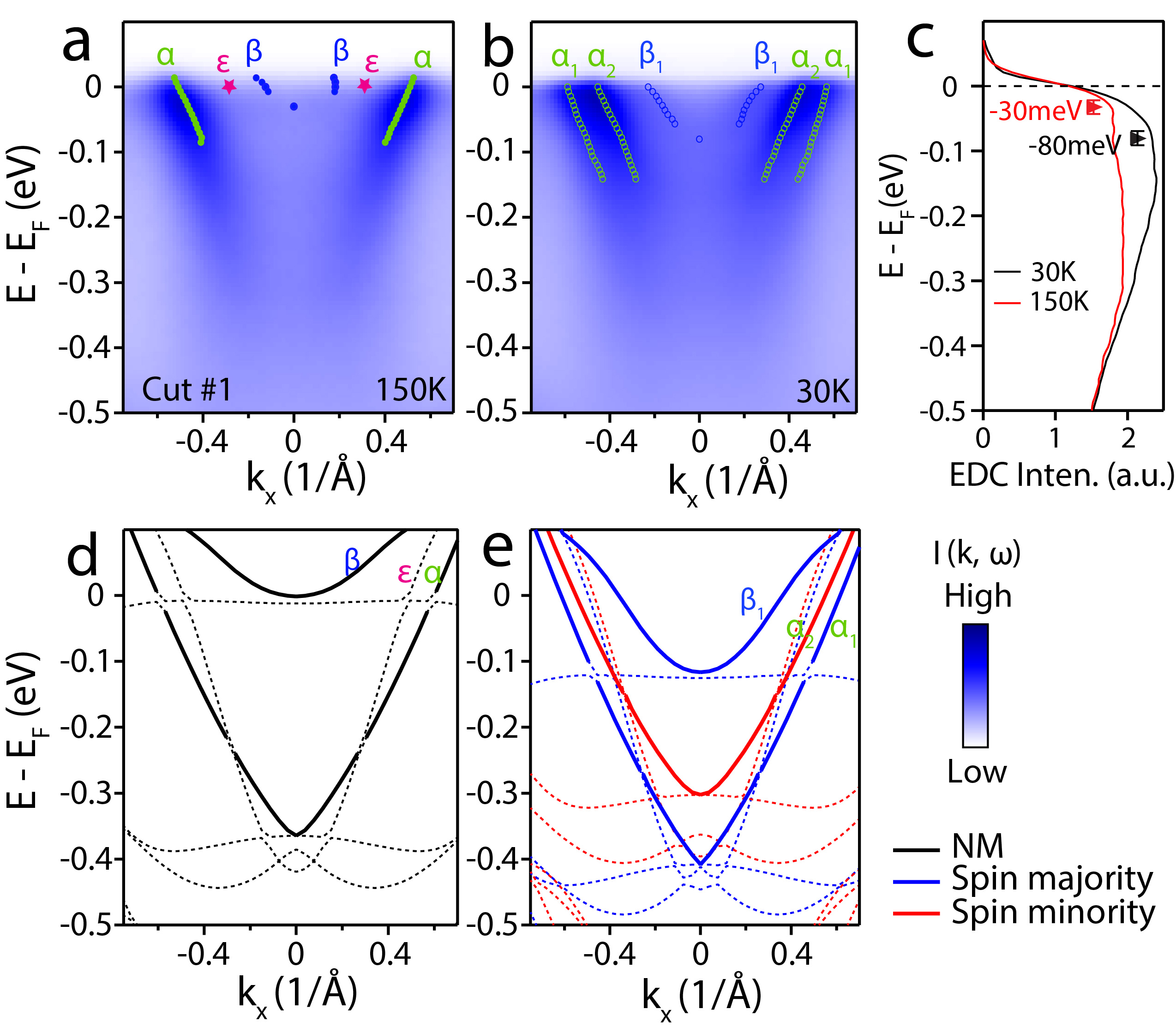}
\caption{\label{fig:Fig3} {\bf Electronic structure across \tc.} (a) Spectral image along cut \#1 in Fig.~\ref{fig:Fig2}a with MDC-fitted dispersions. The k$_F$ value of $\varepsilon$ is obtained from fitting the MDC at the equivalent X$_2$-point. (b) Same as (a) except taken at 30 K in the FM state. (c) Energy distribution curves (EDCs) at the X$_1$ point. The markers indicate the $\beta$ band bottom obtained by fitting the corresponding data points in (a) and (b) with a parabolic curve. Fitting uncertainty gives a $\pm$10 meV error bar. (d) DFT band calculations in the paramagnetic phase renormalized by a factor of 2.9. (e) Same as (d) except in the FM state.}
\end{figure}

The spin-splitting of bands in the presence of the FM order can be further visualized from the band dispersions. Band images measured across X$_1$ in the paramagnetic phase (150 K) are shown in Fig.~\ref{fig:Fig3}a. Related band dispersions obtained by MDC fitting as well as the k$_F$ positions of the $\varepsilon$ electron pocket are overlaid on the image. To understand better the observed bands, we have performed DFT calculations of non-magnetic and FM states of RbCo$_2$Se$_2$. Focusing around the X point in the paramagnetic phase (Fig.~\ref{fig:Fig3}d), our observed dispersions show reasonable comparison with the calculated electron bands, where a total of 3 electron bands appear near \ef. From the size of the \kf~openings, the outermost band and the innermost band (solid black bands) likely correspond to the $\alpha$ and $\beta$ bands observed at X$_1$, while the middle band (dotted black band) correspond to the $\varepsilon$ band observed at X$_2$ (Fig.~\ref{fig:Fig2}a). This assignment is further supported by a consideration of the orbital characters of these bands and the photoemission matrix element effects as will be discussed shortly. To match with the Fermi velocity of the outer-most $\alpha$ band, a renormalization factor of 2.9 is applied to the DFT calculations, suggesting moderate electronic correlations in RbCo$_2$Se$_2$. We also note that there exists an orbital-dependent relative shift that would be needed to match the calculations. Such behavior has been commonly observed in iron-based superconductors. \cite{Yi2009,Brouet2013,Watson2015}

In the FM state (Fig.~\ref{fig:Fig3}b), the $\alpha$ band is observed to split into two bands, i.e. $\alpha_1$ and $\alpha_2$. The $\beta$ band is observed to shift down in energy compared to that in the paramagnetic phase. In the calculated band structure in the FM state (Fig.~\ref{fig:Fig3}e), the $\alpha$ band splits into the spin majority band and spin minority band, marked by solid blue and red curves. Similarly, the $\beta$ band also splits with the spin majority band shifted down while the spin minority band has shifted to well above \ef. The bottom of the $\beta$ band is observed to be at 30 meV below \ef~in the paramagnetic state while the spin majority band is shifted down to 80 meV below \ef~in the FM state (Fig.~\ref{fig:Fig3}c). An estimation based on the assumption of equal spin-splitting therefore locates the spin minority band of $\beta$ to be at 20 meV above \ef, and hence not observed.

\begin{figure}
\centering
\includegraphics[width=0.48\textwidth]{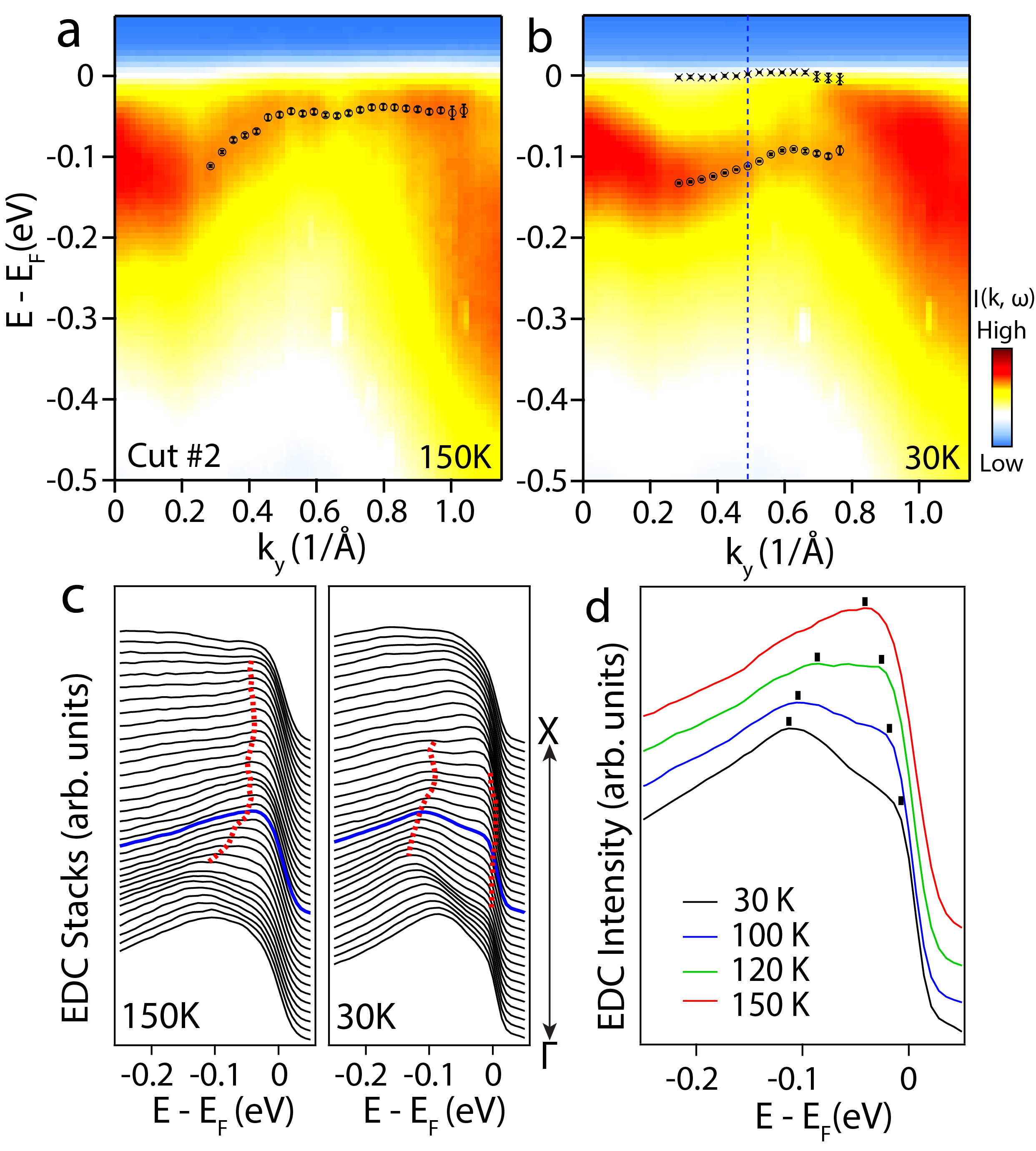}
\caption{\label{fig:Fig4}{\bf Flat band across \tc.} (a,b) Spectra images along cut \#2 in Fig.~\ref{fig:Fig2}a at 150 and 30 K, respectively. Markers are the corresponding fitted peak positions in (c). (c) EDC stacks of (a) and (b). Red markers indicated peak positions from fitting the EDCs with two Lorentzians and a constant background multiplied by the Fermi-Dirac function. (d) Temperature dependent EDCs at the momentum indicated by the blue dashed line in (b).}
\end{figure}

Furthermore, we observe a nearly flat band from the spectra image at cut \#2 along the \gx~direction (Fig.~\ref{fig:Fig4}a). The flat band sits close to \ef~at 150 K in the paramagnetic state. It splits into two at 30 K in the FM state as is evident in the comparison of the EDC stacks between 30 K and 150 K where a single peak at 150 K splits into a peak at lower energy with a shoulder near \ef~at 30 K, which is likely the residual tail of the spin-minority band that has shifted to above \ef~(Fig.~\ref{fig:Fig4}c). The band-splitting behavior of this flat band is reminiscent of that of the $\beta$ band where the spin-majority band is shifted down below \ef~while the spin-minority band shifts to above \ef. The diminishing of the flatband splitting with increasing temperature is shown by EDCs at different temperatures in Fig.~\ref{fig:Fig4}d.

\begin{figure}
\centering
\includegraphics[width=0.48\textwidth]{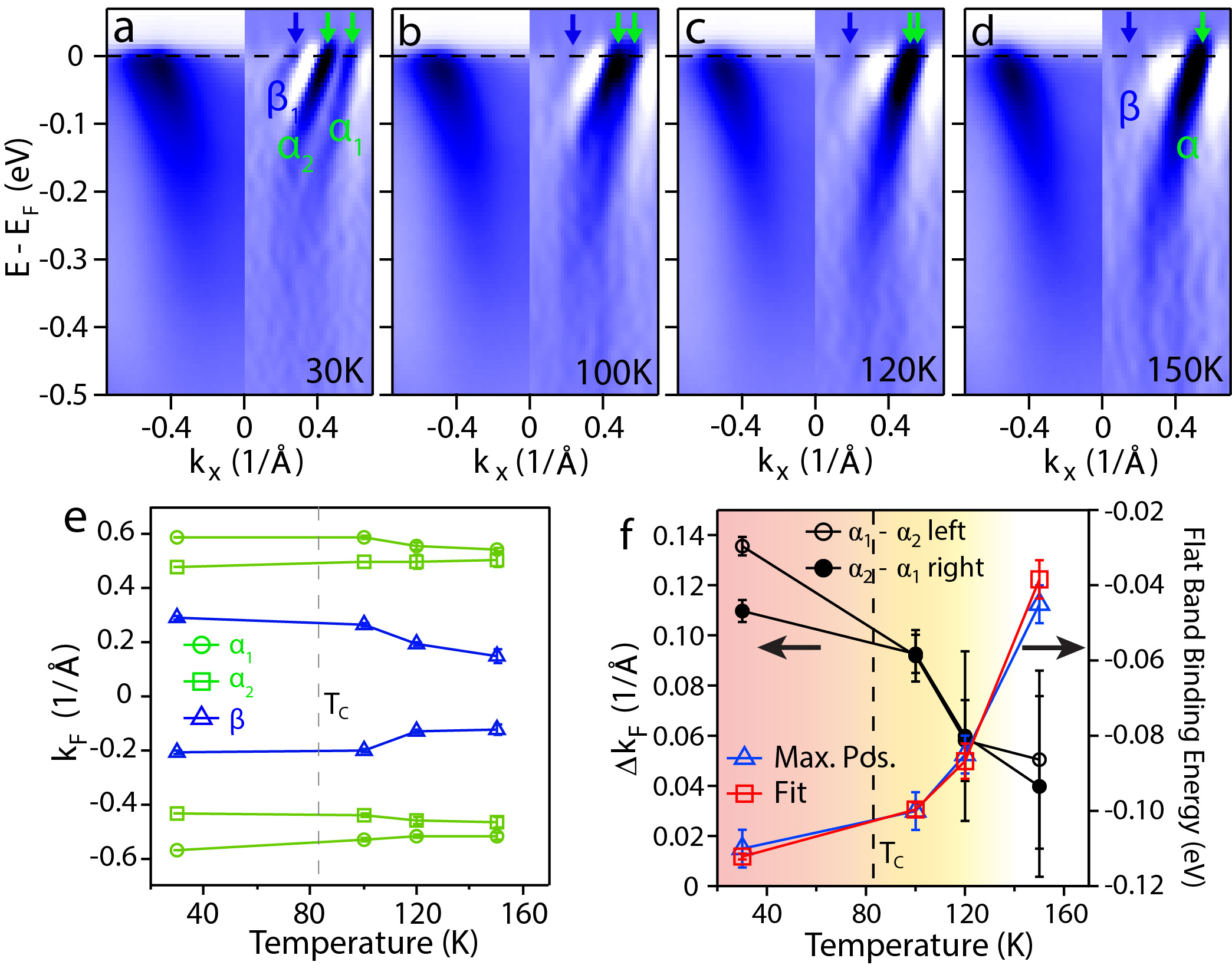}
\caption{\label{fig:Fig5}{\bf Temperature dependence of band splittings.} (a-d) Temperature-dependent spectra along cut \#1 marked in Fig.~\ref{fig:Fig2}a. The left (right) half shows the raw data (second momentum derivative). (e) Fitted \kf~of $\alpha_1$, $\alpha_2$ and $\beta_1$ bands as a function of temperature. Error bars are obtained from the fitting uncertainty. (f) Temperature dependence of $\alpha$ band $\Delta k_F$ and the shift of the flatband from fitting the peak position in the EDCs and maximum positions in Fig.~\ref{fig:Fig4}d}
\end{figure}

To confirm further the relation between the band splitting and the FM order, we examine the evolution of the spin splitting with temperature. With increasing temperature, the $\alpha_1$ and $\alpha_2$ bands are observed to merge (Fig.~\ref{fig:Fig5}a-d). However, they remain split at 100 K. To quantify the splitting against temperature, we extract the \kf~of the $\alpha$ and $\beta$ bands by fitting the MDC at \ef~(Fig.~\ref{fig:Fig5}e), from which we can also extract the \kf~differences (Fig.~\ref{fig:Fig5}f). In addition, the splitting of the flatband can also be tracked from the peaks in the EDCs (Fig.~\ref{fig:Fig4}d). This shows that while spin splitting of bands occurs across \tc, the onset of the splittings persists to a higher temperature than \tc, suggestive of ferromagnetic fluctuations in the paramagnetic state.

\begin{figure}
\includegraphics[width=0.48\textwidth]{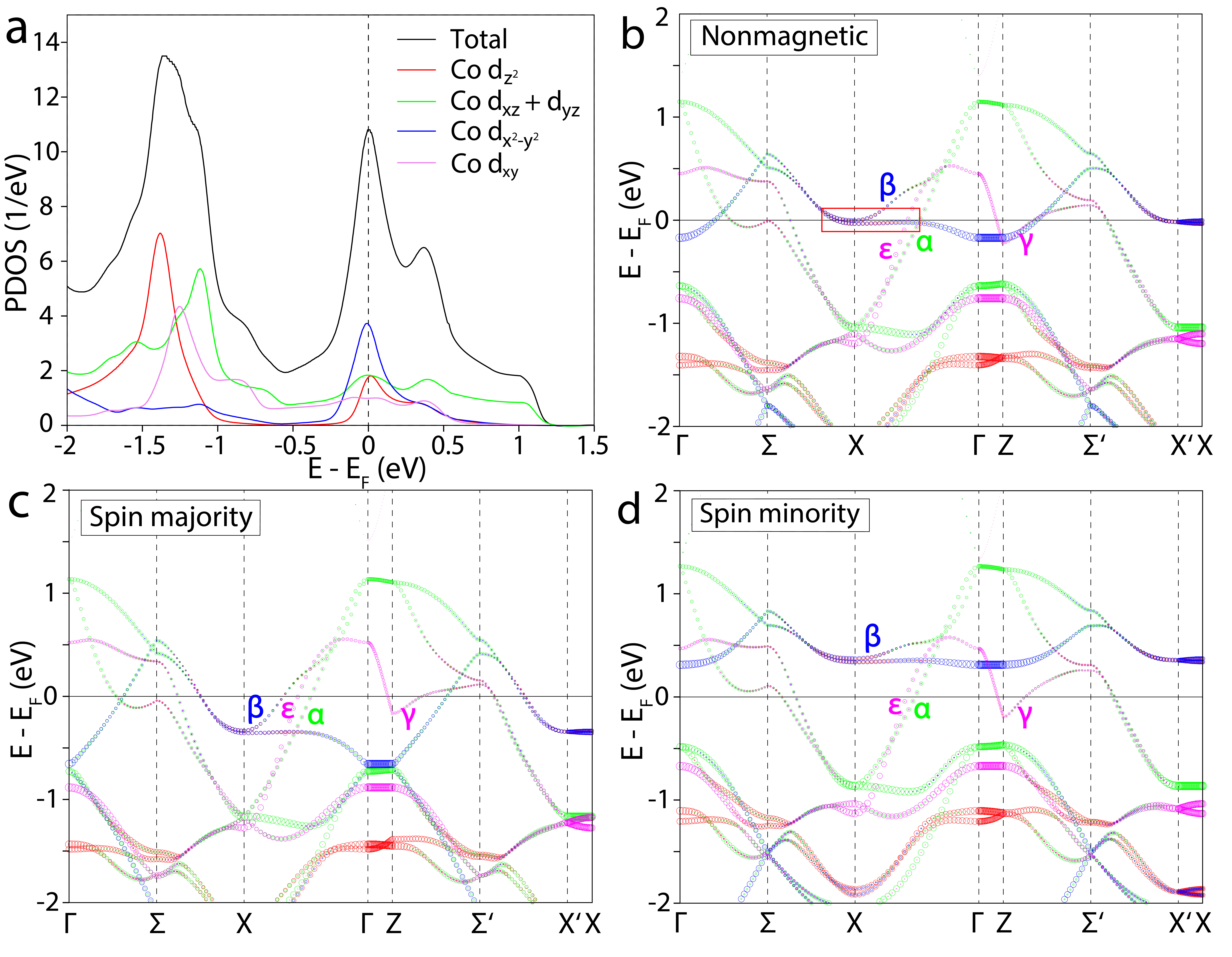}
\caption{\label{fig:Fig6} {\bf Calculated Electronic Structures.} (a) Calculated PDOS projected unto Co $d$ orbitals in the non-magnetic state. (b-d) Calculated orbital-resolved band dispersion in the non-magnetic, spin-majority, and spin-minority state, respectively. The marker size represents the spectral weight.}
\end{figure}

To understand better the behavior of the band-dependent splitting, we carried out orbital-projected DFT calculations of both non-magnetic and FM RbCo$_2$Se$_2$. From the nonmagnetic calculations, three electron bands appear around the X point (Fig.~\ref{fig:Fig6}b), the relative sizes of which allow us to identify the observed $\alpha$, $\beta$ and $\varepsilon$ bands as the calculated bands with dominantly $d_{xz/yz}$, $d_{x^2-y^2}$ and $d_{xy}$ orbital characters, respectively. This assignment is consistent with the expected photoemission matrix element effects, where under the s-polarization used, the $\alpha$ (\dxz/\dyz) band shows strong intensity at the 
%and $\varepsilon$ bands are the dominantly \dxz/\dyz~and \dxy~electron bands at the BZ corner observed in almost all FeSCs. Under s-polarization, the \dxz/\dyz~$\alpha$ band is expected to show strong intensity at the 
X$_1$ point due to its odd symmetry with respect to the \g-X direction, while the $\varepsilon$ (\dxy) band exhibits observable but weaker intensity at the X$_2$ point due to switched parity under the glide mirror symmetry~\cite{Brouet2012}. The innermost $\beta$ band is usually not observable in FeSC parent compounds due to its high kinetic energy, but is now observable due to the heavy electron-doping from Co. We find a qualitative match between the calculated and measured dispersions with a renormalization factor of 2.9 to match the Fermi velocity of the observed $\alpha$ band (Fig.~\ref{fig:Fig3}a). We also note that the $\gamma$~band is a highly \kz-dispersive band where its band bottom is below \ef~near Z and rises to above \ef~near $\Gamma$. Since we do observe the $\gamma$ electron band at the BZ center, we are likely measuring near \kz~= $\pi$. Importantly, a flatband appears at \ef~in the nonmagnetic calculation along \gx~with dominant $d_{x^2-y^2}$ orbital character, consistent with our ARPES measurements (Fig.~\ref{fig:Fig4}a). Such a flatband leads to a large density of states (DOS) at \ef, which could induce a strong FS instability resulting in a band splitting - the spin majority band and minority band - to reduce the overall energy of the system~\cite{EdmundCliftonStoner1938, Blundell2003}.

\begin{table}[b]
\caption{\label{tab:table1}%
Orbital-resolved contribution of the magnetic moment from the Co 3d orbitals, with a total of 0.60(4) $\mu_B$ per Co from all orbitals. }
\begin{ruledtabular}
\begin{tabular}{ccccc}
Orbital&$d_{x^2-y^2}$&$d_{xz/yz}$&$d_{xy}$&$d_{z^2}$\\
\hline
Magnetic moment ($\mu_B$)&0.30&0.10&0.02&0.11\\
\end{tabular}
\end{ruledtabular}
\end{table}

Indeed, a large DOS at \ef~is contributed by this flatband as seen from the calculated orbital-projected DOS for the nonmagnetic phase (Fig.~\ref{fig:Fig6}a). In the calculation for the FM state, the spin splitting between the majority and minority bands is orbital-dependent: largest in the $d_{x^2-y^2}$ flatband, followed by the $d_{xz/yz}$ bands, and finally the $d_{xy}$ bands (Fig.~\ref{fig:Fig6}c and d). This is consistent with our ARPES measurements in that the splittings of the flatband and the $\alpha$ and $\beta$ bands are clearly observed while those of the $\varepsilon$ band are not. Similarly, the $\gamma$ electron band observed at \g~is also dominantly of $d_{xy}$ character, where no significant modification across \tc~is observed (Fig.~\ref{fig:Fig2}a, c). We can confirm further the role of the $d_{x^2-y^2}$ flatband to the itinerant FM by calculating the contribution to the ordered moment from different orbitals (TABLE~\ref{tab:table1}). Indeed, the $d_{x^2-y^2}$ orbital contributes 0.3 $\mu_B$ out of the 0.60(4) $\mu_B$ total ordered magnetic moment per Co site (TABLE~\ref{tab:table1}) while the $d_{xy}$ orbital contributes merely 0.02 $\mu_B$. Our combined experimental and theoretical results are in support of a flatband-induced itinerant origin for the ferromagnetism in RbCo$_2$Se$_2$.

\begin{figure}
\includegraphics[width=0.48\textwidth]{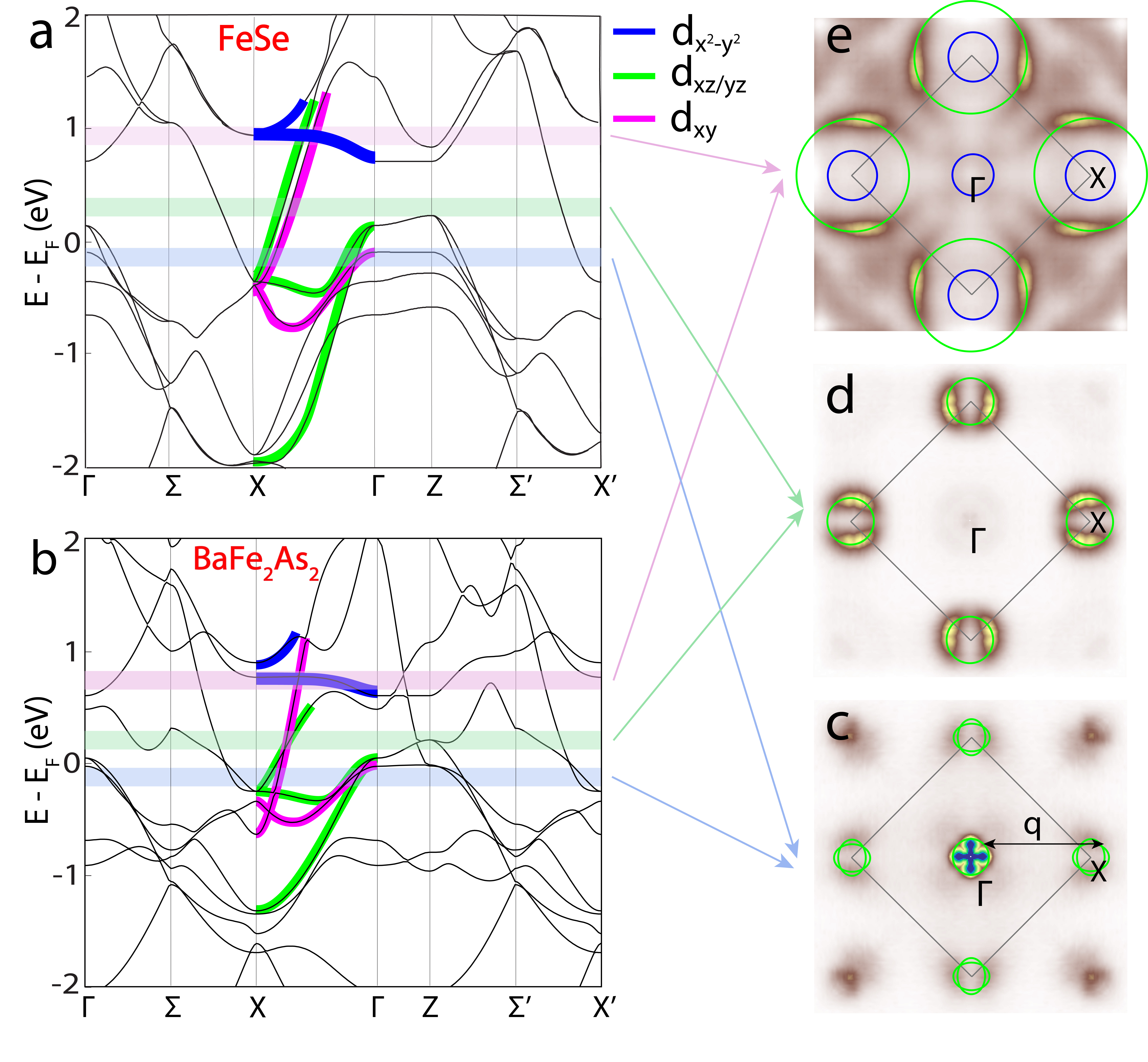}
\caption{\label{fig:Fig7} {\bf Electronic phases in iron-based superconductors.} (a,b) Band structure calculations of FeSe~\cite{Subedi2008} and BaFe$_2$As$_2$~\cite{Graser2010}. Common features along \gx~are highlighted by dominant orbitals. Horizontal lines represent different chemical potentials tunable via electron doping as exemplified by FSs of (c) Fe(Se,Te), (d) high \tc~RbFe$_2$Se$_2$, and (e) flatband itinerant FM RbCo$_2$Se$_2$.}
\end{figure}

Finally, it is interesting to note that this $d_{x^2-y^2}$ flatband is a commonality in the electronic structure of FeSCs. Besides the ACo$_2$Se$_2$ family, SrCo$_2$As$_2$ has also been recently identified to host itinerant ferromagnetism due to a flatband near \ef~\cite{Shen2019a, Li2019}. Taking the observed electronic phases together amongst FeSCs, we point out the following phenomenology in relation to the common electronic structure of the Fe $3d$ orbitals.
%In principle, the physical properties of different quantum materials are manifested by their electronic structures which are determined by the crystal structures. RbCo$_2$Se$_2$ is isostructural with RbFe$_2$Se$_2$. The main difference between them is the electron doping level. How do their physical properties differ completely with RbFe$_2$Se$_2$ being a superconductor\cite{Ivanovskii2011} and RbCo$_2$Se$_2$ exhibiting ferromagnetic ground state\cite{HUAN1989}? 
We use band structure calculations for FeSe~\cite{Subedi2008} and BaFe$_2$As$_2$~\cite{Graser2010} to represent the FeSe-based and FeAs-based materials where common features are highlighted (Fig.~\ref{fig:Fig7}a and b). We illustrate three main types of fermiologies observed. For the undoped parent compounds such as BaFe$_2$As$_2$, NaFeAs, LaFeAsO, and Fe(Te,Se), the chemical potential leads to quasi-nested hole pockets at the BZ center and electron pockets at the BZ corner, where a collinear spin density wave appears at the nesting wavevector with the associated nematic order (Fig.~\ref{fig:Fig7}c)~\cite{Chu2010, Chu2012, Fernandes2014,Yi2009, Liu2009a, Nakayama2014, Watson2015}. Superconductivity appears when these competing orders are suppressed with doping or chemical pressure. With further heavy electron doping, the chemical potential shifts up (green line in Fig.~\ref{fig:Fig7}a and b) such that the hole bands at \g~sink below \ef, leaving enlarged electron pockets at the BZ corner, largely destroying the nesting condition. The group of materials exhibiting this type of fermiology includes the heavily electron-doped iron-chalcogenides such as AFe$_2$Se$_2$~\cite{Qian2011, Mou2011a, Zhang2011a}, (Li,Fe)OHFeSe ~\cite{Zhao2016a, Niu2015}, and monolayer FeSe film grown on \sto~substrate\cite{Liu2012, Tan2013, Lee2014, Song2019}, exhibiting no competing ordered ground states and superconductivity appears with \tc~of 30 K to 65 K. With further electron doping, as achieved by replacing Fe with Co, the chemical potential can be tuned further up (pink line in Fig.~\ref{fig:Fig7}a and b) where the electron pockets at the zone corner are further enlarged. In this doping regime, a flatband appears near \ef, as has been identified in both SrCo$_2$As$_2$ and ACo$_2$Se$_2$, driving the electronic system into an itinerant ferromagnetic state~\cite{Bannikov2011, Quirinale2013, Liu2015b, Li2019a, Li2019}. It is interesting to point out that optimal superconductivity amongst FeSCs appears in the heavily electron-doped iron-chalcogenides where the Fermiology is farthest away from instabilities leading to competing phases--on one side the quasi-nesting inductive of SDW order, and on the other the flatband inductive of itinerant ferromagnetism.
%As ferromagnetism is detrimental to the electron pairing, superconductivity has not been found in these materials\cite{HUAN1989, Shatruk2019}.

\section{CONCLUSION}
To summarize, we report the evolution of the electronic structure of the itinerant ferromagnetic compound RbCo$_2$Se$_2$ across its ferromagnetic transition. In the paramagnetic state, the much enlarged electron Fermi pockets around the BZ corner indicate the heavy electron doping from its superconducting isostructural analog RbFe$_2$Se$_2$. A renormalization factor of 2.9 suggests moderate electron-electron correlations, comparable with those in the iron-based superconductors. We find an orbital-dependent band splitting in the ferromagnetic state. In comparison to first-principle calculations, we find our observations to be consistent with a flatband itinerant origin of the ferromagnetism. Furthermore, the band splitting is observed to persist within a finite temperature window above the ferromagnetic transition, suggesting ferromagnetic fluctuations above the long range ordering temperature. Finally, we point out a phenomenological observation of the appearance of high temperature superconductivity in the iron-based superconductors bounded by competing phases of SDW and nematicity on one side in a quasi-nested Fermiology and flatband itinerant ferromagnetism on the other side, tunable via carrier doping.

%\begin{thebibliography}{10}
%\end{thebibliography}

\begin{acknowledgments} 
This research used resources of the Advanced Light Source, a U.S. DOE Office of Science User Facility under contract no. DE-AC02-05CH11231, and the High-Flux Isotope Reactor, a DOE Office of Science User Facility operated by the Oak Ridge National Laboratory.
%ARPES experiments were performed at the Advanced Light Source, which is operated by the Office of Basic Energy Sciences, U.S. DOE. Work at University of California, Berkeley and Lawrence Berkeley National Laboratory was funded by the U.S. Department of Energy, Office of Science, Office of Basic Energy Sciences, Materials Sciences and Engineering Division under Contract No. DE-AC02-05-CH11231 within the Quantum Materials Program (KC2202) and the Office of Basic Energy Sciences. 
The work at LBL was funded by the U.S. Department of Energy, Office of Science, Office of Basic Energy Sciences, Materials Sciences and Engineering Division under Contract No. DE-AC02-05-CH11231 (Quantum Materials program KC2202). The work at Rice University was supported by the Robert A. Welch Foundation Grant No. C-2024 as well as the Alfred P. Sloan Foundation. Work at Sun Yat-Sen University was supported by the National Natural Science Foundation of China (Grant No. 11904414), Natural Science Foundation of Guangdong  (No. 2018A030313055), National Key Research and Development Program of China (No. 2019YFA0705700), and Young Zhujiang Scholar program. A.F.K. was supported by NSF DMR-1752713.
\end{acknowledgments}

%\noindent {\bf Author contributions}\\
%M.Y. proposed and designed the research. M.Y. carried out the ARPES experiment. J.W.H., Z.C.W., H.S.P., H.W. and M.Y. analyzed the data. J.W.H. and M.Y. wrote the paper. All authors participated in discussion and comment on the paper.\\

%\noindent{\bf Additional information}\\
%Correspondence and requests for materials should be addressed to M.Y.\\

%%\noindent {\bf\large Additional information}\\
%\noindent{\bf Competing interests:} 
%The authors declare no competing interests.

%\bibliography{RbCo2Se2_PRL_submit2.bbl}
%\bibliography{RbCo2Se2_PRL_submit}% Produces the bibliography via BibTeX.
\bibliography{RbCo2Se2_Reference}

\begin{thebibliography}{48}%
\makeatletter
\providecommand \@ifxundefined [1]{%
 \@ifx{#1\undefined}
}%
\providecommand \@ifnum [1]{%
 \ifnum #1\expandafter \@firstoftwo
 \else \expandafter \@secondoftwo
 \fi
}%
\providecommand \@ifx [1]{%
 \ifx #1\expandafter \@firstoftwo
 \else \expandafter \@secondoftwo
 \fi
}%
\providecommand \natexlab [1]{#1}%
\providecommand \enquote  [1]{``#1''}%
\providecommand \bibnamefont  [1]{#1}%
\providecommand \bibfnamefont [1]{#1}%
\providecommand \citenamefont [1]{#1}%
\providecommand \href@noop [0]{\@secondoftwo}%
\providecommand \href [0]{\begingroup \@sanitize@url \@href}%
\providecommand \@href[1]{\@@startlink{#1}\@@href}%
\providecommand \@@href[1]{\endgroup#1\@@endlink}%
\providecommand \@sanitize@url [0]{\catcode `\\12\catcode `\$12\catcode
  `\&12\catcode `\#12\catcode `\^12\catcode `\_12\catcode `\%12\relax}%
\providecommand \@@startlink[1]{}%
\providecommand \@@endlink[0]{}%
\providecommand \url  [0]{\begingroup\@sanitize@url \@url }%
\providecommand \@url [1]{\endgroup\@href {#1}{\urlprefix }}%
\providecommand \urlprefix  [0]{URL }%
\providecommand \Eprint [0]{\href }%
\providecommand \doibase [0]{http://dx.doi.org/}%
\providecommand \selectlanguage [0]{\@gobble}%
\providecommand \bibinfo  [0]{\@secondoftwo}%
\providecommand \bibfield  [0]{\@secondoftwo}%
\providecommand \translation [1]{[#1]}%
\providecommand \BibitemOpen [0]{}%
\providecommand \bibitemStop [0]{}%
\providecommand \bibitemNoStop [0]{.\EOS\space}%
\providecommand \EOS [0]{\spacefactor3000\relax}%
\providecommand \BibitemShut  [1]{\csname bibitem#1\endcsname}%
\let\auto@bib@innerbib\@empty
%</preamble>
\bibitem [{\citenamefont {de~la Cruz}\ \emph {et~al.}(2008)\citenamefont {de~la
  Cruz}, \citenamefont {Huang}, \citenamefont {Lynn}, \citenamefont {Li},
  \citenamefont {II}, \citenamefont {Zarestky}, \citenamefont {Mook},
  \citenamefont {Chen}, \citenamefont {Luo}, \citenamefont {Wang},\ and\
  \citenamefont {Dai}}]{DeLaCruz2008}%
  \BibitemOpen
  \bibfield  {author} {\bibinfo {author} {\bibfnamefont {C.}~\bibnamefont
  {de~la Cruz}}, \bibinfo {author} {\bibfnamefont {Q.}~\bibnamefont {Huang}},
  \bibinfo {author} {\bibfnamefont {J.~W.}\ \bibnamefont {Lynn}}, \bibinfo
  {author} {\bibfnamefont {J.}~\bibnamefont {Li}}, \bibinfo {author}
  {\bibfnamefont {W.~R.}\ \bibnamefont {II}}, \bibinfo {author} {\bibfnamefont
  {J.~L.}\ \bibnamefont {Zarestky}}, \bibinfo {author} {\bibfnamefont {H.~A.}\
  \bibnamefont {Mook}}, \bibinfo {author} {\bibfnamefont {G.~F.}\ \bibnamefont
  {Chen}}, \bibinfo {author} {\bibfnamefont {J.~L.}\ \bibnamefont {Luo}},
  \bibinfo {author} {\bibfnamefont {N.~L.}\ \bibnamefont {Wang}}, \ and\
  \bibinfo {author} {\bibfnamefont {P.}~\bibnamefont {Dai}},\ }\href {\doibase
  10.1038/nature07057} {\bibfield  {journal} {\bibinfo  {journal} {Nature}\
  }\textbf {\bibinfo {volume} {453}},\ \bibinfo {pages} {899} (\bibinfo {year}
  {2008})}\BibitemShut {NoStop}%
\bibitem [{\citenamefont {Yi}\ \emph {et~al.}(2017)\citenamefont {Yi},
  \citenamefont {Zhang}, \citenamefont {Shen},\ and\ \citenamefont
  {Lu}}]{Yi2017}%
  \BibitemOpen
  \bibfield  {author} {\bibinfo {author} {\bibfnamefont {M.}~\bibnamefont
  {Yi}}, \bibinfo {author} {\bibfnamefont {Y.}~\bibnamefont {Zhang}}, \bibinfo
  {author} {\bibfnamefont {Z.-X.}\ \bibnamefont {Shen}}, \ and\ \bibinfo
  {author} {\bibfnamefont {D.}~\bibnamefont {Lu}},\ }\href {\doibase
  10.1038/s41535-017-0059-y} {\bibfield  {journal} {\bibinfo  {journal} {npj
  Quantum Materials}\ }\textbf {\bibinfo {volume} {2}},\ \bibinfo {pages} {57}
  (\bibinfo {year} {2017})}\BibitemShut {NoStop}%
\bibitem [{\citenamefont {Kordyuk}(2012)}]{Kordyuk2012}%
  \BibitemOpen
  \bibfield  {author} {\bibinfo {author} {\bibfnamefont {A.~A.}\ \bibnamefont
  {Kordyuk}},\ }\href {\doibase 10.1063/1.4752092} {\bibfield  {journal}
  {\bibinfo  {journal} {Low Temperature Physics}\ }\textbf {\bibinfo {volume}
  {38}},\ \bibinfo {pages} {888} (\bibinfo {year} {2012})}\BibitemShut
  {NoStop}%
\bibitem [{\citenamefont {Dai}(2015)}]{Dai2015}%
  \BibitemOpen
  \bibfield  {author} {\bibinfo {author} {\bibfnamefont {P.}~\bibnamefont
  {Dai}},\ }\href {\doibase 10.1103/RevModPhys.87.855} {\bibfield  {journal}
  {\bibinfo  {journal} {Reviews of Modern Physics}\ }\textbf {\bibinfo {volume}
  {87}},\ \bibinfo {pages} {855} (\bibinfo {year} {2015})}\BibitemShut
  {NoStop}%
\bibitem [{\citenamefont {Wang}\ \emph {et~al.}(2015)\citenamefont {Wang},
  \citenamefont {Yi}, \citenamefont {Cao}, \citenamefont {de~la Cruz},
  \citenamefont {Mo}, \citenamefont {Huang}, \citenamefont
  {Bourret-Courchesne}, \citenamefont {Dai}, \citenamefont {Lee}, \citenamefont
  {Shen},\ and\ \citenamefont {Birgeneau}}]{Wang2015a}%
  \BibitemOpen
  \bibfield  {author} {\bibinfo {author} {\bibfnamefont {M.}~\bibnamefont
  {Wang}}, \bibinfo {author} {\bibfnamefont {M.}~\bibnamefont {Yi}}, \bibinfo
  {author} {\bibfnamefont {H.}~\bibnamefont {Cao}}, \bibinfo {author}
  {\bibfnamefont {C.}~\bibnamefont {de~la Cruz}}, \bibinfo {author}
  {\bibfnamefont {S.~K.}\ \bibnamefont {Mo}}, \bibinfo {author} {\bibfnamefont
  {Q.~Z.}\ \bibnamefont {Huang}}, \bibinfo {author} {\bibfnamefont
  {E.}~\bibnamefont {Bourret-Courchesne}}, \bibinfo {author} {\bibfnamefont
  {P.}~\bibnamefont {Dai}}, \bibinfo {author} {\bibfnamefont {D.~H.}\
  \bibnamefont {Lee}}, \bibinfo {author} {\bibfnamefont {Z.~X.}\ \bibnamefont
  {Shen}}, \ and\ \bibinfo {author} {\bibfnamefont {R.~J.}\ \bibnamefont
  {Birgeneau}},\ }\href {\doibase 10.1103/PhysRevB.92.121101} {\bibfield
  {journal} {\bibinfo  {journal} {Physical Review B}\ }\textbf {\bibinfo
  {volume} {92}},\ \bibinfo {pages} {121101} (\bibinfo {year}
  {2015})}\BibitemShut {NoStop}%
\bibitem [{\citenamefont {Wang}\ \emph {et~al.}(2016)\citenamefont {Wang},
  \citenamefont {Yi}, \citenamefont {Tian}, \citenamefont
  {Bourret-Courchesne},\ and\ \citenamefont {Birgeneau}}]{Wang2016}%
  \BibitemOpen
  \bibfield  {author} {\bibinfo {author} {\bibfnamefont {M.}~\bibnamefont
  {Wang}}, \bibinfo {author} {\bibfnamefont {M.}~\bibnamefont {Yi}}, \bibinfo
  {author} {\bibfnamefont {W.}~\bibnamefont {Tian}}, \bibinfo {author}
  {\bibfnamefont {E.}~\bibnamefont {Bourret-Courchesne}}, \ and\ \bibinfo
  {author} {\bibfnamefont {R.~J.}\ \bibnamefont {Birgeneau}},\ }\href {\doibase
  10.1103/PhysRevB.93.075155} {\bibfield  {journal} {\bibinfo  {journal}
  {Physical Review B}\ }\textbf {\bibinfo {volume} {93}},\ \bibinfo {pages}
  {075155} (\bibinfo {year} {2016})}\BibitemShut {NoStop}%
\bibitem [{\citenamefont {Huan}\ \emph {et~al.}(1989)\citenamefont {Huan},
  \citenamefont {Greenblatt},\ and\ \citenamefont {Croft}}]{HUAN1989}%
  \BibitemOpen
  \bibfield  {author} {\bibinfo {author} {\bibfnamefont {G.}~\bibnamefont
  {Huan}}, \bibinfo {author} {\bibfnamefont {M.}~\bibnamefont {Greenblatt}}, \
  and\ \bibinfo {author} {\bibfnamefont {M.}~\bibnamefont {Croft}},\ }\href
  {http://doi.wiley.com/10.1002/chin.198950023} {\bibfield  {journal} {\bibinfo
   {journal} {Eur. J. Solid State Inorg. Chem.}\ }\textbf {\bibinfo {volume}
  {26}},\ \bibinfo {pages} {193} (\bibinfo {year} {1989})}\BibitemShut
  {NoStop}%
\bibitem [{\citenamefont {Huan}\ and\ \citenamefont
  {Greenblatt}(1989)}]{Metals1989}%
  \BibitemOpen
  \bibfield  {author} {\bibinfo {author} {\bibfnamefont {G.}~\bibnamefont
  {Huan}}\ and\ \bibinfo {author} {\bibfnamefont {M.}~\bibnamefont
  {Greenblatt}},\ }\href@noop {} {\bibfield  {journal} {\bibinfo  {journal}
  {Journal of the Less-Common Metals}\ }\textbf {\bibinfo {volume} {156}},\
  \bibinfo {pages} {247} (\bibinfo {year} {1989})}\BibitemShut {NoStop}%
\bibitem [{\citenamefont {Yang}\ \emph {et~al.}(2013)\citenamefont {Yang},
  \citenamefont {Chen}, \citenamefont {Wang}, \citenamefont {Mao},
  \citenamefont {Imai}, \citenamefont {Yoshimura},\ and\ \citenamefont
  {Fang}}]{Yang2013}%
  \BibitemOpen
  \bibfield  {author} {\bibinfo {author} {\bibfnamefont {J.}~\bibnamefont
  {Yang}}, \bibinfo {author} {\bibfnamefont {B.}~\bibnamefont {Chen}}, \bibinfo
  {author} {\bibfnamefont {H.}~\bibnamefont {Wang}}, \bibinfo {author}
  {\bibfnamefont {Q.}~\bibnamefont {Mao}}, \bibinfo {author} {\bibfnamefont
  {M.}~\bibnamefont {Imai}}, \bibinfo {author} {\bibfnamefont {K.}~\bibnamefont
  {Yoshimura}}, \ and\ \bibinfo {author} {\bibfnamefont {M.}~\bibnamefont
  {Fang}},\ }\href {\doibase 10.1103/PhysRevB.88.064406} {\bibfield  {journal}
  {\bibinfo  {journal} {Physical Review B}\ }\textbf {\bibinfo {volume} {88}},\
  \bibinfo {pages} {064406} (\bibinfo {year} {2013})}\BibitemShut {NoStop}%
\bibitem [{\citenamefont {von Rohr}\ \emph {et~al.}(2016)\citenamefont {von
  Rohr}, \citenamefont {Krzton-Maziopa}, \citenamefont {Pomjakushin},
  \citenamefont {Grundmann}, \citenamefont {Guguchia}, \citenamefont
  {Schnick},\ and\ \citenamefont {Schilling}}]{VonRohr2016}%
  \BibitemOpen
  \bibfield  {author} {\bibinfo {author} {\bibfnamefont {F.}~\bibnamefont {von
  Rohr}}, \bibinfo {author} {\bibfnamefont {A.}~\bibnamefont {Krzton-Maziopa}},
  \bibinfo {author} {\bibfnamefont {V.}~\bibnamefont {Pomjakushin}}, \bibinfo
  {author} {\bibfnamefont {H.}~\bibnamefont {Grundmann}}, \bibinfo {author}
  {\bibfnamefont {Z.}~\bibnamefont {Guguchia}}, \bibinfo {author}
  {\bibfnamefont {W.}~\bibnamefont {Schnick}}, \ and\ \bibinfo {author}
  {\bibfnamefont {A.}~\bibnamefont {Schilling}},\ }\href {\doibase
  10.1088/0953-8984/28/27/276001} {\bibfield  {journal} {\bibinfo  {journal}
  {Journal of Physics: Condensed Matter}\ }\textbf {\bibinfo {volume} {28}},\
  \bibinfo {pages} {276001} (\bibinfo {year} {2016})}\BibitemShut {NoStop}%
\bibitem [{\citenamefont {Stoner}(1938)}]{EdmundCliftonStoner1938}%
  \BibitemOpen
  \bibfield  {author} {\bibinfo {author} {\bibfnamefont {E.~C.}\ \bibnamefont
  {Stoner}},\ }\href {\doibase 10.1098/rspa.1938.0066} {\bibfield  {journal}
  {\bibinfo  {journal} {Proceedings of the Royal Society of London. Series A.
  Mathematical and Physical Sciences}\ }\textbf {\bibinfo {volume} {165}},\
  \bibinfo {pages} {372} (\bibinfo {year} {1938})}\BibitemShut {NoStop}%
\bibitem [{\citenamefont {Yang}\ \emph {et~al.}(2004)\citenamefont {Yang},
  \citenamefont {Wang}, \citenamefont {Sekharan}, \citenamefont {Matsui},
  \citenamefont {Souma}, \citenamefont {Sato}, \citenamefont {Takahashi},
  \citenamefont {Takeuchi}, \citenamefont {Campuzano}, \citenamefont {Jin},
  \citenamefont {Sales}, \citenamefont {Mandrus}, \citenamefont {Wang},\ and\
  \citenamefont {Ding}}]{Yang2004}%
  \BibitemOpen
  \bibfield  {author} {\bibinfo {author} {\bibfnamefont {H.-B.}\ \bibnamefont
  {Yang}}, \bibinfo {author} {\bibfnamefont {S.-C.}\ \bibnamefont {Wang}},
  \bibinfo {author} {\bibfnamefont {A.~K.~P.}\ \bibnamefont {Sekharan}},
  \bibinfo {author} {\bibfnamefont {H.}~\bibnamefont {Matsui}}, \bibinfo
  {author} {\bibfnamefont {S.}~\bibnamefont {Souma}}, \bibinfo {author}
  {\bibfnamefont {T.}~\bibnamefont {Sato}}, \bibinfo {author} {\bibfnamefont
  {T.}~\bibnamefont {Takahashi}}, \bibinfo {author} {\bibfnamefont
  {T.}~\bibnamefont {Takeuchi}}, \bibinfo {author} {\bibfnamefont {J.~C.}\
  \bibnamefont {Campuzano}}, \bibinfo {author} {\bibfnamefont {R.}~\bibnamefont
  {Jin}}, \bibinfo {author} {\bibfnamefont {B.~C.}\ \bibnamefont {Sales}},
  \bibinfo {author} {\bibfnamefont {D.}~\bibnamefont {Mandrus}}, \bibinfo
  {author} {\bibfnamefont {Z.}~\bibnamefont {Wang}}, \ and\ \bibinfo {author}
  {\bibfnamefont {H.}~\bibnamefont {Ding}},\ }\href {\doibase
  10.1103/PhysRevLett.92.246403} {\bibfield  {journal} {\bibinfo  {journal}
  {Physical Review Letters}\ }\textbf {\bibinfo {volume} {92}},\ \bibinfo
  {pages} {246403} (\bibinfo {year} {2004})}\BibitemShut {NoStop}%
\bibitem [{\citenamefont {G{\"{u}}ttler}\ \emph {et~al.}(2016)\citenamefont
  {G{\"{u}}ttler}, \citenamefont {Generalov}, \citenamefont {Otrokov},
  \citenamefont {Kummer}, \citenamefont {Kliemt}, \citenamefont {Fedorov},
  \citenamefont {Chikina}, \citenamefont {Danzenb{\"{a}}cher}, \citenamefont
  {Schulz}, \citenamefont {Chulkov}, \citenamefont {Koroteev}, \citenamefont
  {Caroca-Canales}, \citenamefont {Shi}, \citenamefont {Radovic}, \citenamefont
  {Geibel}, \citenamefont {Laubschat}, \citenamefont {Dudin}, \citenamefont
  {Kim}, \citenamefont {Hoesch}, \citenamefont {Krellner},\ and\ \citenamefont
  {Vyalikh}}]{Guttler2016}%
  \BibitemOpen
  \bibfield  {author} {\bibinfo {author} {\bibfnamefont {M.}~\bibnamefont
  {G{\"{u}}ttler}}, \bibinfo {author} {\bibfnamefont {A.}~\bibnamefont
  {Generalov}}, \bibinfo {author} {\bibfnamefont {M.~M.}\ \bibnamefont
  {Otrokov}}, \bibinfo {author} {\bibfnamefont {K.}~\bibnamefont {Kummer}},
  \bibinfo {author} {\bibfnamefont {K.}~\bibnamefont {Kliemt}}, \bibinfo
  {author} {\bibfnamefont {A.}~\bibnamefont {Fedorov}}, \bibinfo {author}
  {\bibfnamefont {A.}~\bibnamefont {Chikina}}, \bibinfo {author} {\bibfnamefont
  {S.}~\bibnamefont {Danzenb{\"{a}}cher}}, \bibinfo {author} {\bibfnamefont
  {S.}~\bibnamefont {Schulz}}, \bibinfo {author} {\bibfnamefont {E.~V.}\
  \bibnamefont {Chulkov}}, \bibinfo {author} {\bibfnamefont {Y.~M.}\
  \bibnamefont {Koroteev}}, \bibinfo {author} {\bibfnamefont {N.}~\bibnamefont
  {Caroca-Canales}}, \bibinfo {author} {\bibfnamefont {M.}~\bibnamefont {Shi}},
  \bibinfo {author} {\bibfnamefont {M.}~\bibnamefont {Radovic}}, \bibinfo
  {author} {\bibfnamefont {C.}~\bibnamefont {Geibel}}, \bibinfo {author}
  {\bibfnamefont {C.}~\bibnamefont {Laubschat}}, \bibinfo {author}
  {\bibfnamefont {P.}~\bibnamefont {Dudin}}, \bibinfo {author} {\bibfnamefont
  {T.~K.}\ \bibnamefont {Kim}}, \bibinfo {author} {\bibfnamefont
  {M.}~\bibnamefont {Hoesch}}, \bibinfo {author} {\bibfnamefont
  {C.}~\bibnamefont {Krellner}}, \ and\ \bibinfo {author} {\bibfnamefont
  {D.~V.}\ \bibnamefont {Vyalikh}},\ }\href {\doibase 10.1038/srep24254}
  {\bibfield  {journal} {\bibinfo  {journal} {Scientific Reports}\ }\textbf
  {\bibinfo {volume} {6}},\ \bibinfo {pages} {24254} (\bibinfo {year}
  {2016})}\BibitemShut {NoStop}%
\bibitem [{\citenamefont {Mazzola}\ \emph {et~al.}(2018)\citenamefont
  {Mazzola}, \citenamefont {Sunko}, \citenamefont {Khim}, \citenamefont
  {Rosner}, \citenamefont {Kushwaha}, \citenamefont {Clark}, \citenamefont
  {Bawden}, \citenamefont {Markovi{\'{c}}}, \citenamefont {Kim}, \citenamefont
  {Hoesch}, \citenamefont {Mackenzie},\ and\ \citenamefont
  {King}}]{Mazzola2018}%
  \BibitemOpen
  \bibfield  {author} {\bibinfo {author} {\bibfnamefont {F.}~\bibnamefont
  {Mazzola}}, \bibinfo {author} {\bibfnamefont {V.}~\bibnamefont {Sunko}},
  \bibinfo {author} {\bibfnamefont {S.}~\bibnamefont {Khim}}, \bibinfo {author}
  {\bibfnamefont {H.}~\bibnamefont {Rosner}}, \bibinfo {author} {\bibfnamefont
  {P.}~\bibnamefont {Kushwaha}}, \bibinfo {author} {\bibfnamefont {O.~J.}\
  \bibnamefont {Clark}}, \bibinfo {author} {\bibfnamefont {L.}~\bibnamefont
  {Bawden}}, \bibinfo {author} {\bibfnamefont {I.}~\bibnamefont
  {Markovi{\'{c}}}}, \bibinfo {author} {\bibfnamefont {T.~K.}\ \bibnamefont
  {Kim}}, \bibinfo {author} {\bibfnamefont {M.}~\bibnamefont {Hoesch}},
  \bibinfo {author} {\bibfnamefont {A.~P.}\ \bibnamefont {Mackenzie}}, \ and\
  \bibinfo {author} {\bibfnamefont {P.~D.~C.}\ \bibnamefont {King}},\ }\href
  {\doibase 10.1073/pnas.1811873115} {\bibfield  {journal} {\bibinfo  {journal}
  {Proceedings of the National Academy of Sciences}\ }\textbf {\bibinfo
  {volume} {115}},\ \bibinfo {pages} {12956} (\bibinfo {year}
  {2018})}\BibitemShut {NoStop}%
\bibitem [{\citenamefont {Liu}\ \emph {et~al.}(2015)\citenamefont {Liu},
  \citenamefont {Zhao}, \citenamefont {Li}, \citenamefont {Jia}, \citenamefont
  {Cai}, \citenamefont {Zhou}, \citenamefont {Xia}, \citenamefont
  {B{\"{u}}chner}, \citenamefont {Borisenko},\ and\ \citenamefont
  {Wang}}]{Liu2015b}%
  \BibitemOpen
  \bibfield  {author} {\bibinfo {author} {\bibfnamefont {Z.~H.}\ \bibnamefont
  {Liu}}, \bibinfo {author} {\bibfnamefont {Y.~G.}\ \bibnamefont {Zhao}},
  \bibinfo {author} {\bibfnamefont {Y.}~\bibnamefont {Li}}, \bibinfo {author}
  {\bibfnamefont {L.~L.}\ \bibnamefont {Jia}}, \bibinfo {author} {\bibfnamefont
  {Y.~P.}\ \bibnamefont {Cai}}, \bibinfo {author} {\bibfnamefont
  {S.}~\bibnamefont {Zhou}}, \bibinfo {author} {\bibfnamefont {T.~L.}\
  \bibnamefont {Xia}}, \bibinfo {author} {\bibfnamefont {B.}~\bibnamefont
  {B{\"{u}}chner}}, \bibinfo {author} {\bibfnamefont {S.~V.}\ \bibnamefont
  {Borisenko}}, \ and\ \bibinfo {author} {\bibfnamefont {S.~C.}\ \bibnamefont
  {Wang}},\ }\href {\doibase 10.1088/0953-8984/27/29/295501} {\bibfield
  {journal} {\bibinfo  {journal} {Journal of Physics: Condensed Matter}\
  }\textbf {\bibinfo {volume} {27}},\ \bibinfo {pages} {295501} (\bibinfo
  {year} {2015})}\BibitemShut {NoStop}%
\bibitem [{\citenamefont {Chakoumakos}\ \emph {et~al.}(2011)\citenamefont
  {Chakoumakos}, \citenamefont {Cao}, \citenamefont {Ye}, \citenamefont
  {Stoica}, \citenamefont {Popovici}, \citenamefont {Sundaram}, \citenamefont
  {Zhou}, \citenamefont {Hicks}, \citenamefont {Lynn},\ and\ \citenamefont
  {Riedel}}]{Chakoumakos2011}%
  \BibitemOpen
  \bibfield  {author} {\bibinfo {author} {\bibfnamefont {B.~C.}\ \bibnamefont
  {Chakoumakos}}, \bibinfo {author} {\bibfnamefont {H.}~\bibnamefont {Cao}},
  \bibinfo {author} {\bibfnamefont {F.}~\bibnamefont {Ye}}, \bibinfo {author}
  {\bibfnamefont {A.~D.}\ \bibnamefont {Stoica}}, \bibinfo {author}
  {\bibfnamefont {M.}~\bibnamefont {Popovici}}, \bibinfo {author}
  {\bibfnamefont {M.}~\bibnamefont {Sundaram}}, \bibinfo {author}
  {\bibfnamefont {W.}~\bibnamefont {Zhou}}, \bibinfo {author} {\bibfnamefont
  {J.~S.}\ \bibnamefont {Hicks}}, \bibinfo {author} {\bibfnamefont {G.~W.}\
  \bibnamefont {Lynn}}, \ and\ \bibinfo {author} {\bibfnamefont {R.~A.}\
  \bibnamefont {Riedel}},\ }\href {\doibase 10.1107/S0021889811012301}
  {\bibfield  {journal} {\bibinfo  {journal} {Journal of Applied
  Crystallography}\ }\textbf {\bibinfo {volume} {44}},\ \bibinfo {pages} {655}
  (\bibinfo {year} {2011})}\BibitemShut {NoStop}%
\bibitem [{\citenamefont
  {Rodr{\'{i}}guez-Carvajal}(1993)}]{Rodriguez-Carvajal1993}%
  \BibitemOpen
  \bibfield  {author} {\bibinfo {author} {\bibfnamefont {J.}~\bibnamefont
  {Rodr{\'{i}}guez-Carvajal}},\ }\href {\doibase 10.1016/0921-4526(93)90108-I}
  {\bibfield  {journal} {\bibinfo  {journal} {Physica B: Condensed Matter}\
  }\textbf {\bibinfo {volume} {192}},\ \bibinfo {pages} {55} (\bibinfo {year}
  {1993})}\BibitemShut {NoStop}%
\bibitem [{\citenamefont {Giannozzi}\ \emph {et~al.}(2009)\citenamefont
  {Giannozzi}, \citenamefont {Baroni}, \citenamefont {Bonini}, \citenamefont
  {Calandra}, \citenamefont {Car}, \citenamefont {Cavazzoni}, \citenamefont
  {Ceresoli}, \citenamefont {Chiarotti}, \citenamefont {Cococcioni},
  \citenamefont {Dabo}, \citenamefont {{Dal Corso}}, \citenamefont
  {de~Gironcoli}, \citenamefont {Fabris}, \citenamefont {Fratesi},
  \citenamefont {Gebauer}, \citenamefont {Gerstmann}, \citenamefont
  {Gougoussis}, \citenamefont {Kokalj}, \citenamefont {Lazzeri}, \citenamefont
  {Martin-Samos}, \citenamefont {Marzari}, \citenamefont {Mauri}, \citenamefont
  {Mazzarello}, \citenamefont {Paolini}, \citenamefont {Pasquarello},
  \citenamefont {Paulatto}, \citenamefont {Sbraccia}, \citenamefont {Scandolo},
  \citenamefont {Sclauzero}, \citenamefont {Seitsonen}, \citenamefont
  {Smogunov}, \citenamefont {Umari},\ and\ \citenamefont
  {Wentzcovitch}}]{Giannozzi2009}%
  \BibitemOpen
  \bibfield  {author} {\bibinfo {author} {\bibfnamefont {P.}~\bibnamefont
  {Giannozzi}}, \bibinfo {author} {\bibfnamefont {S.}~\bibnamefont {Baroni}},
  \bibinfo {author} {\bibfnamefont {N.}~\bibnamefont {Bonini}}, \bibinfo
  {author} {\bibfnamefont {M.}~\bibnamefont {Calandra}}, \bibinfo {author}
  {\bibfnamefont {R.}~\bibnamefont {Car}}, \bibinfo {author} {\bibfnamefont
  {C.}~\bibnamefont {Cavazzoni}}, \bibinfo {author} {\bibfnamefont
  {D.}~\bibnamefont {Ceresoli}}, \bibinfo {author} {\bibfnamefont {G.~L.}\
  \bibnamefont {Chiarotti}}, \bibinfo {author} {\bibfnamefont {M.}~\bibnamefont
  {Cococcioni}}, \bibinfo {author} {\bibfnamefont {I.}~\bibnamefont {Dabo}},
  \bibinfo {author} {\bibfnamefont {A.}~\bibnamefont {{Dal Corso}}}, \bibinfo
  {author} {\bibfnamefont {S.}~\bibnamefont {de~Gironcoli}}, \bibinfo {author}
  {\bibfnamefont {S.}~\bibnamefont {Fabris}}, \bibinfo {author} {\bibfnamefont
  {G.}~\bibnamefont {Fratesi}}, \bibinfo {author} {\bibfnamefont
  {R.}~\bibnamefont {Gebauer}}, \bibinfo {author} {\bibfnamefont
  {U.}~\bibnamefont {Gerstmann}}, \bibinfo {author} {\bibfnamefont
  {C.}~\bibnamefont {Gougoussis}}, \bibinfo {author} {\bibfnamefont
  {A.}~\bibnamefont {Kokalj}}, \bibinfo {author} {\bibfnamefont
  {M.}~\bibnamefont {Lazzeri}}, \bibinfo {author} {\bibfnamefont
  {L.}~\bibnamefont {Martin-Samos}}, \bibinfo {author} {\bibfnamefont
  {N.}~\bibnamefont {Marzari}}, \bibinfo {author} {\bibfnamefont
  {F.}~\bibnamefont {Mauri}}, \bibinfo {author} {\bibfnamefont
  {R.}~\bibnamefont {Mazzarello}}, \bibinfo {author} {\bibfnamefont
  {S.}~\bibnamefont {Paolini}}, \bibinfo {author} {\bibfnamefont
  {A.}~\bibnamefont {Pasquarello}}, \bibinfo {author} {\bibfnamefont
  {L.}~\bibnamefont {Paulatto}}, \bibinfo {author} {\bibfnamefont
  {C.}~\bibnamefont {Sbraccia}}, \bibinfo {author} {\bibfnamefont
  {S.}~\bibnamefont {Scandolo}}, \bibinfo {author} {\bibfnamefont
  {G.}~\bibnamefont {Sclauzero}}, \bibinfo {author} {\bibfnamefont {A.~P.}\
  \bibnamefont {Seitsonen}}, \bibinfo {author} {\bibfnamefont {A.}~\bibnamefont
  {Smogunov}}, \bibinfo {author} {\bibfnamefont {P.}~\bibnamefont {Umari}}, \
  and\ \bibinfo {author} {\bibfnamefont {R.~M.}\ \bibnamefont {Wentzcovitch}},\
  }\href {\doibase 10.1088/0953-8984/21/39/395502} {\bibfield  {journal}
  {\bibinfo  {journal} {Journal of Physics: Condensed Matter}\ }\textbf
  {\bibinfo {volume} {21}},\ \bibinfo {pages} {395502} (\bibinfo {year}
  {2009})}\BibitemShut {NoStop}%
\bibitem [{\citenamefont {Giannozzi}\ \emph {et~al.}(2017)\citenamefont
  {Giannozzi}, \citenamefont {Andreussi}, \citenamefont {Brumme}, \citenamefont
  {Bunau}, \citenamefont {{Buongiorno Nardelli}}, \citenamefont {Calandra},
  \citenamefont {Car}, \citenamefont {Cavazzoni}, \citenamefont {Ceresoli},
  \citenamefont {Cococcioni}, \citenamefont {Colonna}, \citenamefont
  {Carnimeo}, \citenamefont {{Dal Corso}}, \citenamefont {{De Gironcoli}},
  \citenamefont {Delugas}, \citenamefont {Distasio}, \citenamefont {Ferretti},
  \citenamefont {Floris}, \citenamefont {Fratesi}, \citenamefont {Fugallo},
  \citenamefont {Gebauer}, \citenamefont {Gerstmann}, \citenamefont {Giustino},
  \citenamefont {Gorni}, \citenamefont {Jia}, \citenamefont {Kawamura},
  \citenamefont {Ko}, \citenamefont {Kokalj}, \citenamefont
  {K{\"{u}}c{\"{u}}kbenli}, \citenamefont {Lazzeri}, \citenamefont {Marsili},
  \citenamefont {Marzari}, \citenamefont {Mauri}, \citenamefont {Nguyen},
  \citenamefont {Nguyen}, \citenamefont {Otero-De-La-Roza}, \citenamefont
  {Paulatto}, \citenamefont {Ponc{\'{e}}}, \citenamefont {Rocca}, \citenamefont
  {Sabatini}, \citenamefont {Santra}, \citenamefont {Schlipf}, \citenamefont
  {Seitsonen}, \citenamefont {Smogunov}, \citenamefont {Timrov}, \citenamefont
  {Thonhauser}, \citenamefont {Umari}, \citenamefont {Vast}, \citenamefont
  {Wu},\ and\ \citenamefont {Baroni}}]{Giannozzi2017}%
  \BibitemOpen
  \bibfield  {author} {\bibinfo {author} {\bibfnamefont {P.}~\bibnamefont
  {Giannozzi}}, \bibinfo {author} {\bibfnamefont {O.}~\bibnamefont
  {Andreussi}}, \bibinfo {author} {\bibfnamefont {T.}~\bibnamefont {Brumme}},
  \bibinfo {author} {\bibfnamefont {O.}~\bibnamefont {Bunau}}, \bibinfo
  {author} {\bibfnamefont {M.}~\bibnamefont {{Buongiorno Nardelli}}}, \bibinfo
  {author} {\bibfnamefont {M.}~\bibnamefont {Calandra}}, \bibinfo {author}
  {\bibfnamefont {R.}~\bibnamefont {Car}}, \bibinfo {author} {\bibfnamefont
  {C.}~\bibnamefont {Cavazzoni}}, \bibinfo {author} {\bibfnamefont
  {D.}~\bibnamefont {Ceresoli}}, \bibinfo {author} {\bibfnamefont
  {M.}~\bibnamefont {Cococcioni}}, \bibinfo {author} {\bibfnamefont
  {N.}~\bibnamefont {Colonna}}, \bibinfo {author} {\bibfnamefont
  {I.}~\bibnamefont {Carnimeo}}, \bibinfo {author} {\bibfnamefont
  {A.}~\bibnamefont {{Dal Corso}}}, \bibinfo {author} {\bibfnamefont
  {S.}~\bibnamefont {{De Gironcoli}}}, \bibinfo {author} {\bibfnamefont
  {P.}~\bibnamefont {Delugas}}, \bibinfo {author} {\bibfnamefont {R.~A.}\
  \bibnamefont {Distasio}}, \bibinfo {author} {\bibfnamefont {A.}~\bibnamefont
  {Ferretti}}, \bibinfo {author} {\bibfnamefont {A.}~\bibnamefont {Floris}},
  \bibinfo {author} {\bibfnamefont {G.}~\bibnamefont {Fratesi}}, \bibinfo
  {author} {\bibfnamefont {G.}~\bibnamefont {Fugallo}}, \bibinfo {author}
  {\bibfnamefont {R.}~\bibnamefont {Gebauer}}, \bibinfo {author} {\bibfnamefont
  {U.}~\bibnamefont {Gerstmann}}, \bibinfo {author} {\bibfnamefont
  {F.}~\bibnamefont {Giustino}}, \bibinfo {author} {\bibfnamefont
  {T.}~\bibnamefont {Gorni}}, \bibinfo {author} {\bibfnamefont
  {J.}~\bibnamefont {Jia}}, \bibinfo {author} {\bibfnamefont {M.}~\bibnamefont
  {Kawamura}}, \bibinfo {author} {\bibfnamefont {H.~Y.}\ \bibnamefont {Ko}},
  \bibinfo {author} {\bibfnamefont {A.}~\bibnamefont {Kokalj}}, \bibinfo
  {author} {\bibfnamefont {E.}~\bibnamefont {K{\"{u}}c{\"{u}}kbenli}}, \bibinfo
  {author} {\bibfnamefont {M.}~\bibnamefont {Lazzeri}}, \bibinfo {author}
  {\bibfnamefont {M.}~\bibnamefont {Marsili}}, \bibinfo {author} {\bibfnamefont
  {N.}~\bibnamefont {Marzari}}, \bibinfo {author} {\bibfnamefont
  {F.}~\bibnamefont {Mauri}}, \bibinfo {author} {\bibfnamefont {N.~L.}\
  \bibnamefont {Nguyen}}, \bibinfo {author} {\bibfnamefont {H.~V.}\
  \bibnamefont {Nguyen}}, \bibinfo {author} {\bibfnamefont {A.}~\bibnamefont
  {Otero-De-La-Roza}}, \bibinfo {author} {\bibfnamefont {L.}~\bibnamefont
  {Paulatto}}, \bibinfo {author} {\bibfnamefont {S.}~\bibnamefont
  {Ponc{\'{e}}}}, \bibinfo {author} {\bibfnamefont {D.}~\bibnamefont {Rocca}},
  \bibinfo {author} {\bibfnamefont {R.}~\bibnamefont {Sabatini}}, \bibinfo
  {author} {\bibfnamefont {B.}~\bibnamefont {Santra}}, \bibinfo {author}
  {\bibfnamefont {M.}~\bibnamefont {Schlipf}}, \bibinfo {author} {\bibfnamefont
  {A.~P.}\ \bibnamefont {Seitsonen}}, \bibinfo {author} {\bibfnamefont
  {A.}~\bibnamefont {Smogunov}}, \bibinfo {author} {\bibfnamefont
  {I.}~\bibnamefont {Timrov}}, \bibinfo {author} {\bibfnamefont
  {T.}~\bibnamefont {Thonhauser}}, \bibinfo {author} {\bibfnamefont
  {P.}~\bibnamefont {Umari}}, \bibinfo {author} {\bibfnamefont
  {N.}~\bibnamefont {Vast}}, \bibinfo {author} {\bibfnamefont {X.}~\bibnamefont
  {Wu}}, \ and\ \bibinfo {author} {\bibfnamefont {S.}~\bibnamefont {Baroni}},\
  }\href {\doibase 10.1088/1361-648X/aa8f79} {\bibfield  {journal} {\bibinfo
  {journal} {Journal of Physics Condensed Matter}\ }\textbf {\bibinfo {volume}
  {29}},\ \bibinfo {pages} {465901} (\bibinfo {year} {2017})}\BibitemShut
  {NoStop}%
\bibitem [{Pot()}]{PotentialforCalculation}%
  \BibitemOpen
  \href@noop {} {\enquote {\bibinfo {title} {{We used the pseudopotentials
  Rb.pbe-sp-hgh.UPF, Co.pbe-nd-rrkjus.UPF, and Se.pbe-n-rrkjus{\_}psl.0.2.UPF
  from the QUANTUM ESPRESSO pseudopotential data base:
  http://www.quantum-espresso.org/pseudopotentials}},}\ }\BibitemShut {NoStop}%
\bibitem [{\citenamefont {Monkhorst}\ and\ \citenamefont
  {Pack}(1976)}]{Monkhorst1976}%
  \BibitemOpen
  \bibfield  {author} {\bibinfo {author} {\bibfnamefont {H.~J.}\ \bibnamefont
  {Monkhorst}}\ and\ \bibinfo {author} {\bibfnamefont {J.~D.}\ \bibnamefont
  {Pack}},\ }\href {\doibase 10.1103/PhysRevB.13.5188} {\bibfield  {journal}
  {\bibinfo  {journal} {Physical Review B}\ }\textbf {\bibinfo {volume} {13}},\
  \bibinfo {pages} {5188} (\bibinfo {year} {1976})}\BibitemShut {NoStop}%
\bibitem [{\citenamefont {Yi}\ \emph {et~al.}(2011)\citenamefont {Yi},
  \citenamefont {Lu}, \citenamefont {Chu}, \citenamefont {Analytis},
  \citenamefont {Sorini}, \citenamefont {Kemper}, \citenamefont {Moritz},
  \citenamefont {Mo}, \citenamefont {Moore}, \citenamefont {Hashimoto},
  \citenamefont {Lee}, \citenamefont {Hussain}, \citenamefont {Devereaux},
  \citenamefont {Fisher},\ and\ \citenamefont {Shen}}]{Yi2011}%
  \BibitemOpen
  \bibfield  {author} {\bibinfo {author} {\bibfnamefont {M.}~\bibnamefont
  {Yi}}, \bibinfo {author} {\bibfnamefont {D.}~\bibnamefont {Lu}}, \bibinfo
  {author} {\bibfnamefont {J.-H.}\ \bibnamefont {Chu}}, \bibinfo {author}
  {\bibfnamefont {J.~G.}\ \bibnamefont {Analytis}}, \bibinfo {author}
  {\bibfnamefont {A.~P.}\ \bibnamefont {Sorini}}, \bibinfo {author}
  {\bibfnamefont {A.~F.}\ \bibnamefont {Kemper}}, \bibinfo {author}
  {\bibfnamefont {B.}~\bibnamefont {Moritz}}, \bibinfo {author} {\bibfnamefont
  {S.-K.}\ \bibnamefont {Mo}}, \bibinfo {author} {\bibfnamefont {R.~G.}\
  \bibnamefont {Moore}}, \bibinfo {author} {\bibfnamefont {M.}~\bibnamefont
  {Hashimoto}}, \bibinfo {author} {\bibfnamefont {W.-S.}\ \bibnamefont {Lee}},
  \bibinfo {author} {\bibfnamefont {Z.}~\bibnamefont {Hussain}}, \bibinfo
  {author} {\bibfnamefont {T.~P.}\ \bibnamefont {Devereaux}}, \bibinfo {author}
  {\bibfnamefont {I.~R.}\ \bibnamefont {Fisher}}, \ and\ \bibinfo {author}
  {\bibfnamefont {Z.-X.}\ \bibnamefont {Shen}},\ }\href {\doibase
  10.1073/pnas.1015572108} {\bibfield  {journal} {\bibinfo  {journal}
  {Proceedings of the National Academy of Sciences}\ }\textbf {\bibinfo
  {volume} {108}},\ \bibinfo {pages} {6878} (\bibinfo {year}
  {2011})}\BibitemShut {NoStop}%
\bibitem [{\citenamefont {Mou}\ \emph {et~al.}(2011)\citenamefont {Mou},
  \citenamefont {Liu}, \citenamefont {Jia}, \citenamefont {He}, \citenamefont
  {Peng}, \citenamefont {Zhao}, \citenamefont {Yu}, \citenamefont {Liu},
  \citenamefont {He}, \citenamefont {Dong}, \citenamefont {Zhang},
  \citenamefont {Wang}, \citenamefont {Dong}, \citenamefont {Fang},
  \citenamefont {Wang}, \citenamefont {Peng}, \citenamefont {Wang},
  \citenamefont {Zhang}, \citenamefont {Yang}, \citenamefont {Xu},
  \citenamefont {Chen},\ and\ \citenamefont {Zhou}}]{Mou2011a}%
  \BibitemOpen
  \bibfield  {author} {\bibinfo {author} {\bibfnamefont {D.}~\bibnamefont
  {Mou}}, \bibinfo {author} {\bibfnamefont {S.}~\bibnamefont {Liu}}, \bibinfo
  {author} {\bibfnamefont {X.}~\bibnamefont {Jia}}, \bibinfo {author}
  {\bibfnamefont {J.}~\bibnamefont {He}}, \bibinfo {author} {\bibfnamefont
  {Y.}~\bibnamefont {Peng}}, \bibinfo {author} {\bibfnamefont {L.}~\bibnamefont
  {Zhao}}, \bibinfo {author} {\bibfnamefont {L.}~\bibnamefont {Yu}}, \bibinfo
  {author} {\bibfnamefont {G.}~\bibnamefont {Liu}}, \bibinfo {author}
  {\bibfnamefont {S.}~\bibnamefont {He}}, \bibinfo {author} {\bibfnamefont
  {X.}~\bibnamefont {Dong}}, \bibinfo {author} {\bibfnamefont {J.}~\bibnamefont
  {Zhang}}, \bibinfo {author} {\bibfnamefont {H.}~\bibnamefont {Wang}},
  \bibinfo {author} {\bibfnamefont {C.}~\bibnamefont {Dong}}, \bibinfo {author}
  {\bibfnamefont {M.}~\bibnamefont {Fang}}, \bibinfo {author} {\bibfnamefont
  {X.}~\bibnamefont {Wang}}, \bibinfo {author} {\bibfnamefont {Q.}~\bibnamefont
  {Peng}}, \bibinfo {author} {\bibfnamefont {Z.}~\bibnamefont {Wang}}, \bibinfo
  {author} {\bibfnamefont {S.}~\bibnamefont {Zhang}}, \bibinfo {author}
  {\bibfnamefont {F.}~\bibnamefont {Yang}}, \bibinfo {author} {\bibfnamefont
  {Z.}~\bibnamefont {Xu}}, \bibinfo {author} {\bibfnamefont {C.}~\bibnamefont
  {Chen}}, \ and\ \bibinfo {author} {\bibfnamefont {X.~J.}\ \bibnamefont
  {Zhou}},\ }\href {\doibase 10.1103/PhysRevLett.106.107001} {\bibfield
  {journal} {\bibinfo  {journal} {Physical Review Letters}\ }\textbf {\bibinfo
  {volume} {106}},\ \bibinfo {pages} {107001} (\bibinfo {year}
  {2011})}\BibitemShut {NoStop}%
\bibitem [{\citenamefont {Zhang}\ \emph {et~al.}(2011)\citenamefont {Zhang},
  \citenamefont {Yang}, \citenamefont {Xu}, \citenamefont {Ye}, \citenamefont
  {Chen}, \citenamefont {He}, \citenamefont {Xu}, \citenamefont {Jiang},
  \citenamefont {Xie}, \citenamefont {Ying}, \citenamefont {Wang},
  \citenamefont {Chen}, \citenamefont {Hu}, \citenamefont {Matsunami},
  \citenamefont {Kimura},\ and\ \citenamefont {Feng}}]{Zhang2011a}%
  \BibitemOpen
  \bibfield  {author} {\bibinfo {author} {\bibfnamefont {Y.}~\bibnamefont
  {Zhang}}, \bibinfo {author} {\bibfnamefont {L.~X.}\ \bibnamefont {Yang}},
  \bibinfo {author} {\bibfnamefont {M.}~\bibnamefont {Xu}}, \bibinfo {author}
  {\bibfnamefont {Z.~R.}\ \bibnamefont {Ye}}, \bibinfo {author} {\bibfnamefont
  {F.}~\bibnamefont {Chen}}, \bibinfo {author} {\bibfnamefont {C.}~\bibnamefont
  {He}}, \bibinfo {author} {\bibfnamefont {H.~C.}\ \bibnamefont {Xu}}, \bibinfo
  {author} {\bibfnamefont {J.}~\bibnamefont {Jiang}}, \bibinfo {author}
  {\bibfnamefont {B.~P.}\ \bibnamefont {Xie}}, \bibinfo {author} {\bibfnamefont
  {J.~J.}\ \bibnamefont {Ying}}, \bibinfo {author} {\bibfnamefont {X.~F.}\
  \bibnamefont {Wang}}, \bibinfo {author} {\bibfnamefont {X.~H.}\ \bibnamefont
  {Chen}}, \bibinfo {author} {\bibfnamefont {J.~P.}\ \bibnamefont {Hu}},
  \bibinfo {author} {\bibfnamefont {M.}~\bibnamefont {Matsunami}}, \bibinfo
  {author} {\bibfnamefont {S.}~\bibnamefont {Kimura}}, \ and\ \bibinfo {author}
  {\bibfnamefont {D.~L.}\ \bibnamefont {Feng}},\ }\href {\doibase
  10.1038/nmat2981} {\bibfield  {journal} {\bibinfo  {journal} {Nature
  Materials}\ }\textbf {\bibinfo {volume} {10}},\ \bibinfo {pages} {273}
  (\bibinfo {year} {2011})}\BibitemShut {NoStop}%
\bibitem [{\citenamefont {Yi}\ \emph {et~al.}(2009)\citenamefont {Yi},
  \citenamefont {Lu}, \citenamefont {Analytis}, \citenamefont {Chu},
  \citenamefont {Mo}, \citenamefont {He}, \citenamefont {Moore}, \citenamefont
  {Zhou}, \citenamefont {Chen}, \citenamefont {Luo}, \citenamefont {Wang},
  \citenamefont {Hussain}, \citenamefont {Singh}, \citenamefont {Fisher},\ and\
  \citenamefont {Shen}}]{Yi2009}%
  \BibitemOpen
  \bibfield  {author} {\bibinfo {author} {\bibfnamefont {M.}~\bibnamefont
  {Yi}}, \bibinfo {author} {\bibfnamefont {D.~H.}\ \bibnamefont {Lu}}, \bibinfo
  {author} {\bibfnamefont {J.~G.}\ \bibnamefont {Analytis}}, \bibinfo {author}
  {\bibfnamefont {J.-H.}\ \bibnamefont {Chu}}, \bibinfo {author} {\bibfnamefont
  {S.-K.}\ \bibnamefont {Mo}}, \bibinfo {author} {\bibfnamefont {R.-H.}\
  \bibnamefont {He}}, \bibinfo {author} {\bibfnamefont {R.~G.}\ \bibnamefont
  {Moore}}, \bibinfo {author} {\bibfnamefont {X.~J.}\ \bibnamefont {Zhou}},
  \bibinfo {author} {\bibfnamefont {G.~F.}\ \bibnamefont {Chen}}, \bibinfo
  {author} {\bibfnamefont {J.~L.}\ \bibnamefont {Luo}}, \bibinfo {author}
  {\bibfnamefont {N.~L.}\ \bibnamefont {Wang}}, \bibinfo {author}
  {\bibfnamefont {Z.}~\bibnamefont {Hussain}}, \bibinfo {author} {\bibfnamefont
  {D.~J.}\ \bibnamefont {Singh}}, \bibinfo {author} {\bibfnamefont {I.~R.}\
  \bibnamefont {Fisher}}, \ and\ \bibinfo {author} {\bibfnamefont {Z.-X.}\
  \bibnamefont {Shen}},\ }\href {\doibase 10.1103/PhysRevB.80.024515}
  {\bibfield  {journal} {\bibinfo  {journal} {Physical Review B}\ }\textbf
  {\bibinfo {volume} {80}},\ \bibinfo {pages} {024515} (\bibinfo {year}
  {2009})}\BibitemShut {NoStop}%
\bibitem [{\citenamefont {Brouet}\ \emph {et~al.}(2013)\citenamefont {Brouet},
  \citenamefont {Lin}, \citenamefont {Texier}, \citenamefont {Bobroff},
  \citenamefont {Taleb-Ibrahimi}, \citenamefont {{Le F{\`{e}}vre}},
  \citenamefont {Bertran}, \citenamefont {Casula}, \citenamefont {Werner},
  \citenamefont {Biermann}, \citenamefont {Rullier-Albenque}, \citenamefont
  {Forget},\ and\ \citenamefont {Colson}}]{Brouet2013}%
  \BibitemOpen
  \bibfield  {author} {\bibinfo {author} {\bibfnamefont {V.}~\bibnamefont
  {Brouet}}, \bibinfo {author} {\bibfnamefont {P.-H.}\ \bibnamefont {Lin}},
  \bibinfo {author} {\bibfnamefont {Y.}~\bibnamefont {Texier}}, \bibinfo
  {author} {\bibfnamefont {J.}~\bibnamefont {Bobroff}}, \bibinfo {author}
  {\bibfnamefont {A.}~\bibnamefont {Taleb-Ibrahimi}}, \bibinfo {author}
  {\bibfnamefont {P.}~\bibnamefont {{Le F{\`{e}}vre}}}, \bibinfo {author}
  {\bibfnamefont {F.}~\bibnamefont {Bertran}}, \bibinfo {author} {\bibfnamefont
  {M.}~\bibnamefont {Casula}}, \bibinfo {author} {\bibfnamefont
  {P.}~\bibnamefont {Werner}}, \bibinfo {author} {\bibfnamefont
  {S.}~\bibnamefont {Biermann}}, \bibinfo {author} {\bibfnamefont
  {F.}~\bibnamefont {Rullier-Albenque}}, \bibinfo {author} {\bibfnamefont
  {A.}~\bibnamefont {Forget}}, \ and\ \bibinfo {author} {\bibfnamefont
  {D.}~\bibnamefont {Colson}},\ }\href {\doibase
  10.1103/PhysRevLett.110.167002} {\bibfield  {journal} {\bibinfo  {journal}
  {Physical Review Letters}\ }\textbf {\bibinfo {volume} {110}},\ \bibinfo
  {pages} {167002} (\bibinfo {year} {2013})}\BibitemShut {NoStop}%
\bibitem [{\citenamefont {Watson}\ \emph {et~al.}(2015)\citenamefont {Watson},
  \citenamefont {Kim}, \citenamefont {Haghighirad}, \citenamefont {Davies},
  \citenamefont {McCollam}, \citenamefont {Narayanan}, \citenamefont {Blake},
  \citenamefont {Chen}, \citenamefont {Ghannadzadeh}, \citenamefont
  {Schofield}, \citenamefont {Hoesch}, \citenamefont {Meingast}, \citenamefont
  {Wolf},\ and\ \citenamefont {Coldea}}]{Watson2015}%
  \BibitemOpen
  \bibfield  {author} {\bibinfo {author} {\bibfnamefont {M.~D.}\ \bibnamefont
  {Watson}}, \bibinfo {author} {\bibfnamefont {T.~K.}\ \bibnamefont {Kim}},
  \bibinfo {author} {\bibfnamefont {A.~A.}\ \bibnamefont {Haghighirad}},
  \bibinfo {author} {\bibfnamefont {N.~R.}\ \bibnamefont {Davies}}, \bibinfo
  {author} {\bibfnamefont {A.}~\bibnamefont {McCollam}}, \bibinfo {author}
  {\bibfnamefont {A.}~\bibnamefont {Narayanan}}, \bibinfo {author}
  {\bibfnamefont {S.~F.}\ \bibnamefont {Blake}}, \bibinfo {author}
  {\bibfnamefont {Y.~L.}\ \bibnamefont {Chen}}, \bibinfo {author}
  {\bibfnamefont {S.}~\bibnamefont {Ghannadzadeh}}, \bibinfo {author}
  {\bibfnamefont {A.~J.}\ \bibnamefont {Schofield}}, \bibinfo {author}
  {\bibfnamefont {M.}~\bibnamefont {Hoesch}}, \bibinfo {author} {\bibfnamefont
  {C.}~\bibnamefont {Meingast}}, \bibinfo {author} {\bibfnamefont
  {T.}~\bibnamefont {Wolf}}, \ and\ \bibinfo {author} {\bibfnamefont {A.~I.}\
  \bibnamefont {Coldea}},\ }\href {\doibase 10.1103/PhysRevB.91.155106}
  {\bibfield  {journal} {\bibinfo  {journal} {Physical Review B}\ }\textbf
  {\bibinfo {volume} {91}},\ \bibinfo {pages} {155106} (\bibinfo {year}
  {2015})}\BibitemShut {NoStop}%
\bibitem [{\citenamefont {Brouet}\ \emph {et~al.}(2012)\citenamefont {Brouet},
  \citenamefont {Jensen}, \citenamefont {Lin}, \citenamefont {Taleb-Ibrahimi},
  \citenamefont {{Le F{\`{e}}vre}}, \citenamefont {Bertran}, \citenamefont
  {Lin}, \citenamefont {Ku}, \citenamefont {Forget},\ and\ \citenamefont
  {Colson}}]{Brouet2012}%
  \BibitemOpen
  \bibfield  {author} {\bibinfo {author} {\bibfnamefont {V.}~\bibnamefont
  {Brouet}}, \bibinfo {author} {\bibfnamefont {M.~F.}\ \bibnamefont {Jensen}},
  \bibinfo {author} {\bibfnamefont {P.-H.}\ \bibnamefont {Lin}}, \bibinfo
  {author} {\bibfnamefont {A.}~\bibnamefont {Taleb-Ibrahimi}}, \bibinfo
  {author} {\bibfnamefont {P.}~\bibnamefont {{Le F{\`{e}}vre}}}, \bibinfo
  {author} {\bibfnamefont {F.}~\bibnamefont {Bertran}}, \bibinfo {author}
  {\bibfnamefont {C.-H.}\ \bibnamefont {Lin}}, \bibinfo {author} {\bibfnamefont
  {W.}~\bibnamefont {Ku}}, \bibinfo {author} {\bibfnamefont {A.}~\bibnamefont
  {Forget}}, \ and\ \bibinfo {author} {\bibfnamefont {D.}~\bibnamefont
  {Colson}},\ }\href {\doibase 10.1103/PhysRevB.86.075123} {\bibfield
  {journal} {\bibinfo  {journal} {Physical Review B}\ }\textbf {\bibinfo
  {volume} {86}},\ \bibinfo {pages} {075123} (\bibinfo {year}
  {2012})}\BibitemShut {NoStop}%
\bibitem [{\citenamefont {Blundell}\ and\ \citenamefont
  {Thouless}(2003)}]{Blundell2003}%
  \BibitemOpen
  \bibfield  {author} {\bibinfo {author} {\bibfnamefont {S.}~\bibnamefont
  {Blundell}}\ and\ \bibinfo {author} {\bibfnamefont {D.}~\bibnamefont
  {Thouless}},\ }\href {\doibase 10.1119/1.1522704} {\bibfield  {journal}
  {\bibinfo  {journal} {American Journal of Physics}\ }\textbf {\bibinfo
  {volume} {71}},\ \bibinfo {pages} {94} (\bibinfo {year} {2003})}\BibitemShut
  {NoStop}%
\bibitem [{\citenamefont {Subedi}\ \emph {et~al.}(2008)\citenamefont {Subedi},
  \citenamefont {Zhang}, \citenamefont {Singh},\ and\ \citenamefont
  {Du}}]{Subedi2008}%
  \BibitemOpen
  \bibfield  {author} {\bibinfo {author} {\bibfnamefont {A.}~\bibnamefont
  {Subedi}}, \bibinfo {author} {\bibfnamefont {L.}~\bibnamefont {Zhang}},
  \bibinfo {author} {\bibfnamefont {D.~J.}\ \bibnamefont {Singh}}, \ and\
  \bibinfo {author} {\bibfnamefont {M.~H.}\ \bibnamefont {Du}},\ }\href
  {\doibase 10.1103/PhysRevB.78.134514} {\bibfield  {journal} {\bibinfo
  {journal} {Physical Review B}\ }\textbf {\bibinfo {volume} {78}},\ \bibinfo
  {pages} {134514} (\bibinfo {year} {2008})}\BibitemShut {NoStop}%
\bibitem [{\citenamefont {Graser}\ \emph {et~al.}(2010)\citenamefont {Graser},
  \citenamefont {Kemper}, \citenamefont {Maier}, \citenamefont {Cheng},
  \citenamefont {Hirschfeld},\ and\ \citenamefont {Scalapino}}]{Graser2010}%
  \BibitemOpen
  \bibfield  {author} {\bibinfo {author} {\bibfnamefont {S.}~\bibnamefont
  {Graser}}, \bibinfo {author} {\bibfnamefont {A.~F.}\ \bibnamefont {Kemper}},
  \bibinfo {author} {\bibfnamefont {T.~A.}\ \bibnamefont {Maier}}, \bibinfo
  {author} {\bibfnamefont {H.-P.}\ \bibnamefont {Cheng}}, \bibinfo {author}
  {\bibfnamefont {P.~J.}\ \bibnamefont {Hirschfeld}}, \ and\ \bibinfo {author}
  {\bibfnamefont {D.~J.}\ \bibnamefont {Scalapino}},\ }\href {\doibase
  10.1103/PhysRevB.81.214503} {\bibfield  {journal} {\bibinfo  {journal}
  {Physical Review B}\ }\textbf {\bibinfo {volume} {81}},\ \bibinfo {pages}
  {214503} (\bibinfo {year} {2010})}\BibitemShut {NoStop}%
\bibitem [{\citenamefont {Shen}\ \emph {et~al.}(2019)\citenamefont {Shen},
  \citenamefont {Zhong}, \citenamefont {Li}, \citenamefont {Lin}, \citenamefont
  {Wang}, \citenamefont {Gu},\ and\ \citenamefont {Feng}}]{Shen2019a}%
  \BibitemOpen
  \bibfield  {author} {\bibinfo {author} {\bibfnamefont {S.}~\bibnamefont
  {Shen}}, \bibinfo {author} {\bibfnamefont {W.}~\bibnamefont {Zhong}},
  \bibinfo {author} {\bibfnamefont {D.}~\bibnamefont {Li}}, \bibinfo {author}
  {\bibfnamefont {Z.}~\bibnamefont {Lin}}, \bibinfo {author} {\bibfnamefont
  {Z.}~\bibnamefont {Wang}}, \bibinfo {author} {\bibfnamefont {X.}~\bibnamefont
  {Gu}}, \ and\ \bibinfo {author} {\bibfnamefont {S.}~\bibnamefont {Feng}},\
  }\href {\doibase 10.1016/j.inoche.2019.03.010} {\bibfield  {journal}
  {\bibinfo  {journal} {Inorganic Chemistry Communications}\ }\textbf {\bibinfo
  {volume} {103}},\ \bibinfo {pages} {25} (\bibinfo {year} {2019})}\BibitemShut
  {NoStop}%
\bibitem [{\citenamefont {Li}\ \emph {et~al.}(2019{\natexlab{a}})\citenamefont
  {Li}, \citenamefont {Yin}, \citenamefont {Liu}, \citenamefont {Wang},
  \citenamefont {Xu}, \citenamefont {Song}, \citenamefont {Tian}, \citenamefont
  {Huang}, \citenamefont {Shen}, \citenamefont {Abernathy}, \citenamefont
  {Niedziela}, \citenamefont {Ewings}, \citenamefont {Perring}, \citenamefont
  {Pajerowski}, \citenamefont {Matsuda}, \citenamefont {Bourges}, \citenamefont
  {Mechthild}, \citenamefont {Su},\ and\ \citenamefont {Dai}}]{Li2019}%
  \BibitemOpen
  \bibfield  {author} {\bibinfo {author} {\bibfnamefont {Y.}~\bibnamefont
  {Li}}, \bibinfo {author} {\bibfnamefont {Z.}~\bibnamefont {Yin}}, \bibinfo
  {author} {\bibfnamefont {Z.}~\bibnamefont {Liu}}, \bibinfo {author}
  {\bibfnamefont {W.}~\bibnamefont {Wang}}, \bibinfo {author} {\bibfnamefont
  {Z.}~\bibnamefont {Xu}}, \bibinfo {author} {\bibfnamefont {Y.}~\bibnamefont
  {Song}}, \bibinfo {author} {\bibfnamefont {L.}~\bibnamefont {Tian}}, \bibinfo
  {author} {\bibfnamefont {Y.}~\bibnamefont {Huang}}, \bibinfo {author}
  {\bibfnamefont {D.}~\bibnamefont {Shen}}, \bibinfo {author} {\bibfnamefont
  {D.~L.}\ \bibnamefont {Abernathy}}, \bibinfo {author} {\bibfnamefont {J.~L.}\
  \bibnamefont {Niedziela}}, \bibinfo {author} {\bibfnamefont {R.~A.}\
  \bibnamefont {Ewings}}, \bibinfo {author} {\bibfnamefont {T.~G.}\
  \bibnamefont {Perring}}, \bibinfo {author} {\bibfnamefont {D.~M.}\
  \bibnamefont {Pajerowski}}, \bibinfo {author} {\bibfnamefont
  {M.}~\bibnamefont {Matsuda}}, \bibinfo {author} {\bibfnamefont
  {P.}~\bibnamefont {Bourges}}, \bibinfo {author} {\bibfnamefont
  {E.}~\bibnamefont {Mechthild}}, \bibinfo {author} {\bibfnamefont
  {Y.}~\bibnamefont {Su}}, \ and\ \bibinfo {author} {\bibfnamefont
  {P.}~\bibnamefont {Dai}},\ }\href {\doibase 10.1103/PhysRevLett.122.117204}
  {\bibfield  {journal} {\bibinfo  {journal} {Physical Review Letters}\
  }\textbf {\bibinfo {volume} {122}},\ \bibinfo {pages} {117204} (\bibinfo
  {year} {2019}{\natexlab{a}})}\BibitemShut {NoStop}%
\bibitem [{\citenamefont {Chu}\ \emph {et~al.}(2010)\citenamefont {Chu},
  \citenamefont {Analytis}, \citenamefont {{De Greve}}, \citenamefont
  {McMahon}, \citenamefont {Islam}, \citenamefont {Yamamoto},\ and\
  \citenamefont {Fisher}}]{Chu2010}%
  \BibitemOpen
  \bibfield  {author} {\bibinfo {author} {\bibfnamefont {J.-H.}\ \bibnamefont
  {Chu}}, \bibinfo {author} {\bibfnamefont {J.~G.}\ \bibnamefont {Analytis}},
  \bibinfo {author} {\bibfnamefont {K.}~\bibnamefont {{De Greve}}}, \bibinfo
  {author} {\bibfnamefont {P.~L.}\ \bibnamefont {McMahon}}, \bibinfo {author}
  {\bibfnamefont {Z.}~\bibnamefont {Islam}}, \bibinfo {author} {\bibfnamefont
  {Y.}~\bibnamefont {Yamamoto}}, \ and\ \bibinfo {author} {\bibfnamefont
  {I.~R.}\ \bibnamefont {Fisher}},\ }\href {\doibase 10.1126/science.1190482}
  {\bibfield  {journal} {\bibinfo  {journal} {Science}\ }\textbf {\bibinfo
  {volume} {329}},\ \bibinfo {pages} {824} (\bibinfo {year}
  {2010})}\BibitemShut {NoStop}%
\bibitem [{\citenamefont {Chu}\ \emph {et~al.}(2012)\citenamefont {Chu},
  \citenamefont {Kuo}, \citenamefont {Analytis},\ and\ \citenamefont
  {Fisher}}]{Chu2012}%
  \BibitemOpen
  \bibfield  {author} {\bibinfo {author} {\bibfnamefont {J.-H.}\ \bibnamefont
  {Chu}}, \bibinfo {author} {\bibfnamefont {H.-H.}\ \bibnamefont {Kuo}},
  \bibinfo {author} {\bibfnamefont {J.~G.}\ \bibnamefont {Analytis}}, \ and\
  \bibinfo {author} {\bibfnamefont {I.~R.}\ \bibnamefont {Fisher}},\ }\href
  {\doibase 10.1126/science.1221713} {\bibfield  {journal} {\bibinfo  {journal}
  {Science}\ }\textbf {\bibinfo {volume} {337}},\ \bibinfo {pages} {710}
  (\bibinfo {year} {2012})}\BibitemShut {NoStop}%
\bibitem [{\citenamefont {Fernandes}\ \emph {et~al.}(2014)\citenamefont
  {Fernandes}, \citenamefont {Chubukov},\ and\ \citenamefont
  {Schmalian}}]{Fernandes2014}%
  \BibitemOpen
  \bibfield  {author} {\bibinfo {author} {\bibfnamefont {R.~M.}\ \bibnamefont
  {Fernandes}}, \bibinfo {author} {\bibfnamefont {A.~V.}\ \bibnamefont
  {Chubukov}}, \ and\ \bibinfo {author} {\bibfnamefont {J.}~\bibnamefont
  {Schmalian}},\ }\href {\doibase 10.1038/nphys2877} {\bibfield  {journal}
  {\bibinfo  {journal} {Nature Physics}\ }\textbf {\bibinfo {volume} {10}},\
  \bibinfo {pages} {97} (\bibinfo {year} {2014})}\BibitemShut {NoStop}%
\bibitem [{\citenamefont {Liu}\ \emph {et~al.}(2009)\citenamefont {Liu},
  \citenamefont {Liu}, \citenamefont {Zhao}, \citenamefont {Zhang},
  \citenamefont {Jia}, \citenamefont {Meng}, \citenamefont {Dong},
  \citenamefont {Zhang}, \citenamefont {Chen}, \citenamefont {Wang},
  \citenamefont {Zhou}, \citenamefont {Zhu}, \citenamefont {Wang},
  \citenamefont {Xu}, \citenamefont {Chen},\ and\ \citenamefont
  {Zhou}}]{Liu2009a}%
  \BibitemOpen
  \bibfield  {author} {\bibinfo {author} {\bibfnamefont {G.}~\bibnamefont
  {Liu}}, \bibinfo {author} {\bibfnamefont {H.}~\bibnamefont {Liu}}, \bibinfo
  {author} {\bibfnamefont {L.}~\bibnamefont {Zhao}}, \bibinfo {author}
  {\bibfnamefont {W.}~\bibnamefont {Zhang}}, \bibinfo {author} {\bibfnamefont
  {X.}~\bibnamefont {Jia}}, \bibinfo {author} {\bibfnamefont {J.}~\bibnamefont
  {Meng}}, \bibinfo {author} {\bibfnamefont {X.}~\bibnamefont {Dong}}, \bibinfo
  {author} {\bibfnamefont {J.}~\bibnamefont {Zhang}}, \bibinfo {author}
  {\bibfnamefont {G.~F.}\ \bibnamefont {Chen}}, \bibinfo {author}
  {\bibfnamefont {G.}~\bibnamefont {Wang}}, \bibinfo {author} {\bibfnamefont
  {Y.}~\bibnamefont {Zhou}}, \bibinfo {author} {\bibfnamefont {Y.}~\bibnamefont
  {Zhu}}, \bibinfo {author} {\bibfnamefont {X.}~\bibnamefont {Wang}}, \bibinfo
  {author} {\bibfnamefont {Z.}~\bibnamefont {Xu}}, \bibinfo {author}
  {\bibfnamefont {C.}~\bibnamefont {Chen}}, \ and\ \bibinfo {author}
  {\bibfnamefont {X.~J.}\ \bibnamefont {Zhou}},\ }\href {\doibase
  10.1103/PhysRevB.80.134519} {\bibfield  {journal} {\bibinfo  {journal}
  {Physical Review B}\ }\textbf {\bibinfo {volume} {80}},\ \bibinfo {pages}
  {134519} (\bibinfo {year} {2009})}\BibitemShut {NoStop}%
\bibitem [{\citenamefont {Nakayama}\ \emph {et~al.}(2014)\citenamefont
  {Nakayama}, \citenamefont {Miyata}, \citenamefont {Phan}, \citenamefont
  {Sato}, \citenamefont {Tanabe}, \citenamefont {Urata}, \citenamefont
  {Tanigaki},\ and\ \citenamefont {Takahashi}}]{Nakayama2014}%
  \BibitemOpen
  \bibfield  {author} {\bibinfo {author} {\bibfnamefont {K.}~\bibnamefont
  {Nakayama}}, \bibinfo {author} {\bibfnamefont {Y.}~\bibnamefont {Miyata}},
  \bibinfo {author} {\bibfnamefont {G.~N.}\ \bibnamefont {Phan}}, \bibinfo
  {author} {\bibfnamefont {T.}~\bibnamefont {Sato}}, \bibinfo {author}
  {\bibfnamefont {Y.}~\bibnamefont {Tanabe}}, \bibinfo {author} {\bibfnamefont
  {T.}~\bibnamefont {Urata}}, \bibinfo {author} {\bibfnamefont
  {K.}~\bibnamefont {Tanigaki}}, \ and\ \bibinfo {author} {\bibfnamefont
  {T.}~\bibnamefont {Takahashi}},\ }\href {\doibase
  10.1103/PhysRevLett.113.237001} {\bibfield  {journal} {\bibinfo  {journal}
  {Physical Review Letters}\ }\textbf {\bibinfo {volume} {113}},\ \bibinfo
  {pages} {237001} (\bibinfo {year} {2014})}\BibitemShut {NoStop}%
\bibitem [{\citenamefont {Qian}\ \emph {et~al.}(2011)\citenamefont {Qian},
  \citenamefont {Wang}, \citenamefont {Jin}, \citenamefont {Zhang},
  \citenamefont {Richard}, \citenamefont {Xu}, \citenamefont {Dai},
  \citenamefont {Fang}, \citenamefont {Guo}, \citenamefont {Chen},\ and\
  \citenamefont {Ding}}]{Qian2011}%
  \BibitemOpen
  \bibfield  {author} {\bibinfo {author} {\bibfnamefont {T.}~\bibnamefont
  {Qian}}, \bibinfo {author} {\bibfnamefont {X.-P.}\ \bibnamefont {Wang}},
  \bibinfo {author} {\bibfnamefont {W.-C.}\ \bibnamefont {Jin}}, \bibinfo
  {author} {\bibfnamefont {P.}~\bibnamefont {Zhang}}, \bibinfo {author}
  {\bibfnamefont {P.}~\bibnamefont {Richard}}, \bibinfo {author} {\bibfnamefont
  {G.}~\bibnamefont {Xu}}, \bibinfo {author} {\bibfnamefont {X.}~\bibnamefont
  {Dai}}, \bibinfo {author} {\bibfnamefont {Z.}~\bibnamefont {Fang}}, \bibinfo
  {author} {\bibfnamefont {J.-G.}\ \bibnamefont {Guo}}, \bibinfo {author}
  {\bibfnamefont {X.-L.}\ \bibnamefont {Chen}}, \ and\ \bibinfo {author}
  {\bibfnamefont {H.}~\bibnamefont {Ding}},\ }\href {\doibase
  10.1103/PhysRevLett.106.187001} {\bibfield  {journal} {\bibinfo  {journal}
  {Physical Review Letters}\ }\textbf {\bibinfo {volume} {106}},\ \bibinfo
  {pages} {187001} (\bibinfo {year} {2011})}\BibitemShut {NoStop}%
\bibitem [{\citenamefont {Zhao}\ \emph {et~al.}(2016)\citenamefont {Zhao},
  \citenamefont {Liang}, \citenamefont {Yuan}, \citenamefont {Hu},
  \citenamefont {Liu}, \citenamefont {Huang}, \citenamefont {He}, \citenamefont
  {Shen}, \citenamefont {Xu}, \citenamefont {Liu}, \citenamefont {Yu},
  \citenamefont {Liu}, \citenamefont {Zhou}, \citenamefont {Huang},
  \citenamefont {Dong}, \citenamefont {Zhou}, \citenamefont {Liu},
  \citenamefont {Lu}, \citenamefont {Zhao}, \citenamefont {Chen}, \citenamefont
  {Xu},\ and\ \citenamefont {Zhou}}]{Zhao2016a}%
  \BibitemOpen
  \bibfield  {author} {\bibinfo {author} {\bibfnamefont {L.}~\bibnamefont
  {Zhao}}, \bibinfo {author} {\bibfnamefont {A.}~\bibnamefont {Liang}},
  \bibinfo {author} {\bibfnamefont {D.}~\bibnamefont {Yuan}}, \bibinfo {author}
  {\bibfnamefont {Y.}~\bibnamefont {Hu}}, \bibinfo {author} {\bibfnamefont
  {D.}~\bibnamefont {Liu}}, \bibinfo {author} {\bibfnamefont {J.}~\bibnamefont
  {Huang}}, \bibinfo {author} {\bibfnamefont {S.}~\bibnamefont {He}}, \bibinfo
  {author} {\bibfnamefont {B.}~\bibnamefont {Shen}}, \bibinfo {author}
  {\bibfnamefont {Y.}~\bibnamefont {Xu}}, \bibinfo {author} {\bibfnamefont
  {X.}~\bibnamefont {Liu}}, \bibinfo {author} {\bibfnamefont {L.}~\bibnamefont
  {Yu}}, \bibinfo {author} {\bibfnamefont {G.}~\bibnamefont {Liu}}, \bibinfo
  {author} {\bibfnamefont {H.}~\bibnamefont {Zhou}}, \bibinfo {author}
  {\bibfnamefont {Y.}~\bibnamefont {Huang}}, \bibinfo {author} {\bibfnamefont
  {X.}~\bibnamefont {Dong}}, \bibinfo {author} {\bibfnamefont {F.}~\bibnamefont
  {Zhou}}, \bibinfo {author} {\bibfnamefont {K.}~\bibnamefont {Liu}}, \bibinfo
  {author} {\bibfnamefont {Z.}~\bibnamefont {Lu}}, \bibinfo {author}
  {\bibfnamefont {Z.}~\bibnamefont {Zhao}}, \bibinfo {author} {\bibfnamefont
  {C.}~\bibnamefont {Chen}}, \bibinfo {author} {\bibfnamefont {Z.}~\bibnamefont
  {Xu}}, \ and\ \bibinfo {author} {\bibfnamefont {X.~J.}\ \bibnamefont
  {Zhou}},\ }\href {\doibase 10.1038/ncomms10608} {\bibfield  {journal}
  {\bibinfo  {journal} {Nature Communications}\ }\textbf {\bibinfo {volume}
  {7}},\ \bibinfo {pages} {10608} (\bibinfo {year} {2016})}\BibitemShut
  {NoStop}%
\bibitem [{\citenamefont {Niu}\ \emph {et~al.}(2015)\citenamefont {Niu},
  \citenamefont {Peng}, \citenamefont {Xu}, \citenamefont {Yan}, \citenamefont
  {Jiang}, \citenamefont {Xu}, \citenamefont {Yu}, \citenamefont {Song},
  \citenamefont {Huang}, \citenamefont {Wang}, \citenamefont {Xie},
  \citenamefont {Lu}, \citenamefont {Wang}, \citenamefont {Chen}, \citenamefont
  {Sun},\ and\ \citenamefont {Feng}}]{Niu2015}%
  \BibitemOpen
  \bibfield  {author} {\bibinfo {author} {\bibfnamefont {X.~H.}\ \bibnamefont
  {Niu}}, \bibinfo {author} {\bibfnamefont {R.}~\bibnamefont {Peng}}, \bibinfo
  {author} {\bibfnamefont {H.~C.}\ \bibnamefont {Xu}}, \bibinfo {author}
  {\bibfnamefont {Y.~J.}\ \bibnamefont {Yan}}, \bibinfo {author} {\bibfnamefont
  {J.}~\bibnamefont {Jiang}}, \bibinfo {author} {\bibfnamefont {D.~F.}\
  \bibnamefont {Xu}}, \bibinfo {author} {\bibfnamefont {T.~L.}\ \bibnamefont
  {Yu}}, \bibinfo {author} {\bibfnamefont {Q.}~\bibnamefont {Song}}, \bibinfo
  {author} {\bibfnamefont {Z.~C.}\ \bibnamefont {Huang}}, \bibinfo {author}
  {\bibfnamefont {Y.~X.}\ \bibnamefont {Wang}}, \bibinfo {author}
  {\bibfnamefont {B.~P.}\ \bibnamefont {Xie}}, \bibinfo {author} {\bibfnamefont
  {X.~F.}\ \bibnamefont {Lu}}, \bibinfo {author} {\bibfnamefont {N.~Z.}\
  \bibnamefont {Wang}}, \bibinfo {author} {\bibfnamefont {X.~H.}\ \bibnamefont
  {Chen}}, \bibinfo {author} {\bibfnamefont {Z.}~\bibnamefont {Sun}}, \ and\
  \bibinfo {author} {\bibfnamefont {D.~L.}\ \bibnamefont {Feng}},\ }\href
  {\doibase 10.1103/PhysRevB.92.060504} {\bibfield  {journal} {\bibinfo
  {journal} {Physical Review B}\ }\textbf {\bibinfo {volume} {92}},\ \bibinfo
  {pages} {060504} (\bibinfo {year} {2015})}\BibitemShut {NoStop}%
\bibitem [{\citenamefont {Liu}\ \emph {et~al.}(2012)\citenamefont {Liu},
  \citenamefont {Zhang}, \citenamefont {Mou}, \citenamefont {He}, \citenamefont
  {Ou}, \citenamefont {Wang}, \citenamefont {Li}, \citenamefont {Wang},
  \citenamefont {Zhao}, \citenamefont {He}, \citenamefont {Peng}, \citenamefont
  {Liu}, \citenamefont {Chen}, \citenamefont {Yu}, \citenamefont {Liu},
  \citenamefont {Dong}, \citenamefont {Zhang}, \citenamefont {Chen},
  \citenamefont {Xu}, \citenamefont {Hu}, \citenamefont {Chen}, \citenamefont
  {Ma}, \citenamefont {Xue},\ and\ \citenamefont {Zhou}}]{Liu2012}%
  \BibitemOpen
  \bibfield  {author} {\bibinfo {author} {\bibfnamefont {D.}~\bibnamefont
  {Liu}}, \bibinfo {author} {\bibfnamefont {W.}~\bibnamefont {Zhang}}, \bibinfo
  {author} {\bibfnamefont {D.}~\bibnamefont {Mou}}, \bibinfo {author}
  {\bibfnamefont {J.}~\bibnamefont {He}}, \bibinfo {author} {\bibfnamefont
  {Y.-B.}\ \bibnamefont {Ou}}, \bibinfo {author} {\bibfnamefont {Q.-Y.}\
  \bibnamefont {Wang}}, \bibinfo {author} {\bibfnamefont {Z.}~\bibnamefont
  {Li}}, \bibinfo {author} {\bibfnamefont {L.}~\bibnamefont {Wang}}, \bibinfo
  {author} {\bibfnamefont {L.}~\bibnamefont {Zhao}}, \bibinfo {author}
  {\bibfnamefont {S.}~\bibnamefont {He}}, \bibinfo {author} {\bibfnamefont
  {Y.}~\bibnamefont {Peng}}, \bibinfo {author} {\bibfnamefont {X.}~\bibnamefont
  {Liu}}, \bibinfo {author} {\bibfnamefont {C.}~\bibnamefont {Chen}}, \bibinfo
  {author} {\bibfnamefont {L.}~\bibnamefont {Yu}}, \bibinfo {author}
  {\bibfnamefont {G.}~\bibnamefont {Liu}}, \bibinfo {author} {\bibfnamefont
  {X.}~\bibnamefont {Dong}}, \bibinfo {author} {\bibfnamefont {J.}~\bibnamefont
  {Zhang}}, \bibinfo {author} {\bibfnamefont {C.}~\bibnamefont {Chen}},
  \bibinfo {author} {\bibfnamefont {Z.}~\bibnamefont {Xu}}, \bibinfo {author}
  {\bibfnamefont {J.}~\bibnamefont {Hu}}, \bibinfo {author} {\bibfnamefont
  {X.}~\bibnamefont {Chen}}, \bibinfo {author} {\bibfnamefont {X.}~\bibnamefont
  {Ma}}, \bibinfo {author} {\bibfnamefont {Q.}~\bibnamefont {Xue}}, \ and\
  \bibinfo {author} {\bibfnamefont {X.}~\bibnamefont {Zhou}},\ }\href {\doibase
  10.1038/ncomms1946} {\bibfield  {journal} {\bibinfo  {journal} {Nature
  Communications}\ }\textbf {\bibinfo {volume} {3}},\ \bibinfo {pages} {931}
  (\bibinfo {year} {2012})}\BibitemShut {NoStop}%
\bibitem [{\citenamefont {Tan}\ \emph {et~al.}(2013)\citenamefont {Tan},
  \citenamefont {Zhang}, \citenamefont {Xia}, \citenamefont {Ye}, \citenamefont
  {Chen}, \citenamefont {Xie}, \citenamefont {Peng}, \citenamefont {Xu},
  \citenamefont {Fan}, \citenamefont {Xu}, \citenamefont {Jiang}, \citenamefont
  {Zhang}, \citenamefont {Lai}, \citenamefont {Xiang}, \citenamefont {Hu},
  \citenamefont {Xie},\ and\ \citenamefont {Feng}}]{Tan2013}%
  \BibitemOpen
  \bibfield  {author} {\bibinfo {author} {\bibfnamefont {S.}~\bibnamefont
  {Tan}}, \bibinfo {author} {\bibfnamefont {Y.}~\bibnamefont {Zhang}}, \bibinfo
  {author} {\bibfnamefont {M.}~\bibnamefont {Xia}}, \bibinfo {author}
  {\bibfnamefont {Z.}~\bibnamefont {Ye}}, \bibinfo {author} {\bibfnamefont
  {F.}~\bibnamefont {Chen}}, \bibinfo {author} {\bibfnamefont {X.}~\bibnamefont
  {Xie}}, \bibinfo {author} {\bibfnamefont {R.}~\bibnamefont {Peng}}, \bibinfo
  {author} {\bibfnamefont {D.}~\bibnamefont {Xu}}, \bibinfo {author}
  {\bibfnamefont {Q.}~\bibnamefont {Fan}}, \bibinfo {author} {\bibfnamefont
  {H.}~\bibnamefont {Xu}}, \bibinfo {author} {\bibfnamefont {J.}~\bibnamefont
  {Jiang}}, \bibinfo {author} {\bibfnamefont {T.}~\bibnamefont {Zhang}},
  \bibinfo {author} {\bibfnamefont {X.}~\bibnamefont {Lai}}, \bibinfo {author}
  {\bibfnamefont {T.}~\bibnamefont {Xiang}}, \bibinfo {author} {\bibfnamefont
  {J.}~\bibnamefont {Hu}}, \bibinfo {author} {\bibfnamefont {B.}~\bibnamefont
  {Xie}}, \ and\ \bibinfo {author} {\bibfnamefont {D.}~\bibnamefont {Feng}},\
  }\href {\doibase 10.1038/nmat3654} {\bibfield  {journal} {\bibinfo  {journal}
  {Nature Materials}\ }\textbf {\bibinfo {volume} {12}},\ \bibinfo {pages}
  {634} (\bibinfo {year} {2013})}\BibitemShut {NoStop}%
\bibitem [{\citenamefont {Lee}\ \emph {et~al.}(2014)\citenamefont {Lee},
  \citenamefont {Schmitt}, \citenamefont {Moore}, \citenamefont {Johnston},
  \citenamefont {Cui}, \citenamefont {Li}, \citenamefont {Yi}, \citenamefont
  {Liu}, \citenamefont {Hashimoto}, \citenamefont {Zhang}, \citenamefont {Lu},
  \citenamefont {Devereaux}, \citenamefont {Lee},\ and\ \citenamefont
  {Shen}}]{Lee2014}%
  \BibitemOpen
  \bibfield  {author} {\bibinfo {author} {\bibfnamefont {J.~J.}\ \bibnamefont
  {Lee}}, \bibinfo {author} {\bibfnamefont {F.~T.}\ \bibnamefont {Schmitt}},
  \bibinfo {author} {\bibfnamefont {R.~G.}\ \bibnamefont {Moore}}, \bibinfo
  {author} {\bibfnamefont {S.}~\bibnamefont {Johnston}}, \bibinfo {author}
  {\bibfnamefont {Y.-T.}\ \bibnamefont {Cui}}, \bibinfo {author} {\bibfnamefont
  {W.}~\bibnamefont {Li}}, \bibinfo {author} {\bibfnamefont {M.}~\bibnamefont
  {Yi}}, \bibinfo {author} {\bibfnamefont {Z.~K.}\ \bibnamefont {Liu}},
  \bibinfo {author} {\bibfnamefont {M.}~\bibnamefont {Hashimoto}}, \bibinfo
  {author} {\bibfnamefont {Y.}~\bibnamefont {Zhang}}, \bibinfo {author}
  {\bibfnamefont {D.~H.}\ \bibnamefont {Lu}}, \bibinfo {author} {\bibfnamefont
  {T.~P.}\ \bibnamefont {Devereaux}}, \bibinfo {author} {\bibfnamefont {D.-H.}\
  \bibnamefont {Lee}}, \ and\ \bibinfo {author} {\bibfnamefont {Z.-X.}\
  \bibnamefont {Shen}},\ }\href {\doibase 10.1038/nature13894} {\bibfield
  {journal} {\bibinfo  {journal} {Nature}\ }\textbf {\bibinfo {volume} {515}},\
  \bibinfo {pages} {245} (\bibinfo {year} {2014})}\BibitemShut {NoStop}%
\bibitem [{\citenamefont {Song}\ \emph {et~al.}(2019)\citenamefont {Song},
  \citenamefont {Yu}, \citenamefont {Lou}, \citenamefont {Xie}, \citenamefont
  {Xu}, \citenamefont {Wen}, \citenamefont {Yao}, \citenamefont {Zhang},
  \citenamefont {Zhu}, \citenamefont {Guo}, \citenamefont {Peng},\ and\
  \citenamefont {Feng}}]{Song2019}%
  \BibitemOpen
  \bibfield  {author} {\bibinfo {author} {\bibfnamefont {Q.}~\bibnamefont
  {Song}}, \bibinfo {author} {\bibfnamefont {T.~L.}\ \bibnamefont {Yu}},
  \bibinfo {author} {\bibfnamefont {X.}~\bibnamefont {Lou}}, \bibinfo {author}
  {\bibfnamefont {B.~P.}\ \bibnamefont {Xie}}, \bibinfo {author} {\bibfnamefont
  {H.~C.}\ \bibnamefont {Xu}}, \bibinfo {author} {\bibfnamefont {C.~H.~P.}\
  \bibnamefont {Wen}}, \bibinfo {author} {\bibfnamefont {Q.}~\bibnamefont
  {Yao}}, \bibinfo {author} {\bibfnamefont {S.~Y.}\ \bibnamefont {Zhang}},
  \bibinfo {author} {\bibfnamefont {X.~T.}\ \bibnamefont {Zhu}}, \bibinfo
  {author} {\bibfnamefont {J.~D.}\ \bibnamefont {Guo}}, \bibinfo {author}
  {\bibfnamefont {R.}~\bibnamefont {Peng}}, \ and\ \bibinfo {author}
  {\bibfnamefont {D.~L.}\ \bibnamefont {Feng}},\ }\href {\doibase
  10.1038/s41467-019-08560-z} {\bibfield  {journal} {\bibinfo  {journal}
  {Nature Communications}\ }\textbf {\bibinfo {volume} {10}},\ \bibinfo {pages}
  {758} (\bibinfo {year} {2019})}\BibitemShut {NoStop}%
\bibitem [{\citenamefont {Bannikov}\ \emph {et~al.}(2012)\citenamefont
  {Bannikov}, \citenamefont {Shein},\ and\ \citenamefont
  {Ivanovskii}}]{Bannikov2011}%
  \BibitemOpen
  \bibfield  {author} {\bibinfo {author} {\bibfnamefont {V.}~\bibnamefont
  {Bannikov}}, \bibinfo {author} {\bibfnamefont {I.}~\bibnamefont {Shein}}, \
  and\ \bibinfo {author} {\bibfnamefont {A.}~\bibnamefont {Ivanovskii}},\
  }\href {\doibase 10.1016/j.physb.2011.10.046} {\bibfield  {journal} {\bibinfo
   {journal} {Physica B: Condensed Matter}\ }\textbf {\bibinfo {volume}
  {407}},\ \bibinfo {pages} {271} (\bibinfo {year} {2012})}\BibitemShut
  {NoStop}%
\bibitem [{\citenamefont {Quirinale}\ \emph {et~al.}(2013)\citenamefont
  {Quirinale}, \citenamefont {Anand}, \citenamefont {Kim}, \citenamefont
  {Pandey}, \citenamefont {Huq}, \citenamefont {Stephens}, \citenamefont
  {Heitmann}, \citenamefont {Kreyssig}, \citenamefont {McQueeney},
  \citenamefont {Johnston},\ and\ \citenamefont {Goldman}}]{Quirinale2013}%
  \BibitemOpen
  \bibfield  {author} {\bibinfo {author} {\bibfnamefont {D.~G.}\ \bibnamefont
  {Quirinale}}, \bibinfo {author} {\bibfnamefont {V.~K.}\ \bibnamefont
  {Anand}}, \bibinfo {author} {\bibfnamefont {M.~G.}\ \bibnamefont {Kim}},
  \bibinfo {author} {\bibfnamefont {A.}~\bibnamefont {Pandey}}, \bibinfo
  {author} {\bibfnamefont {A.}~\bibnamefont {Huq}}, \bibinfo {author}
  {\bibfnamefont {P.~W.}\ \bibnamefont {Stephens}}, \bibinfo {author}
  {\bibfnamefont {T.~W.}\ \bibnamefont {Heitmann}}, \bibinfo {author}
  {\bibfnamefont {A.}~\bibnamefont {Kreyssig}}, \bibinfo {author}
  {\bibfnamefont {R.~J.}\ \bibnamefont {McQueeney}}, \bibinfo {author}
  {\bibfnamefont {D.~C.}\ \bibnamefont {Johnston}}, \ and\ \bibinfo {author}
  {\bibfnamefont {A.~I.}\ \bibnamefont {Goldman}},\ }\href {\doibase
  10.1103/PhysRevB.88.174420} {\bibfield  {journal} {\bibinfo  {journal}
  {Physical Review B}\ }\textbf {\bibinfo {volume} {88}},\ \bibinfo {pages}
  {174420} (\bibinfo {year} {2013})}\BibitemShut {NoStop}%
\bibitem [{\citenamefont {Li}\ \emph {et~al.}(2019{\natexlab{b}})\citenamefont
  {Li}, \citenamefont {Liu}, \citenamefont {Xu}, \citenamefont {Song},
  \citenamefont {Huang}, \citenamefont {Shen}, \citenamefont {Ma},
  \citenamefont {Li}, \citenamefont {Chi}, \citenamefont {Frontzek},
  \citenamefont {Cao}, \citenamefont {Huang}, \citenamefont {Wang},
  \citenamefont {Xie}, \citenamefont {Zhang}, \citenamefont {Rong},
  \citenamefont {Shelton}, \citenamefont {Young}, \citenamefont {DiTusa},\ and\
  \citenamefont {Dai}}]{Li2019a}%
  \BibitemOpen
  \bibfield  {author} {\bibinfo {author} {\bibfnamefont {Y.}~\bibnamefont
  {Li}}, \bibinfo {author} {\bibfnamefont {Z.}~\bibnamefont {Liu}}, \bibinfo
  {author} {\bibfnamefont {Z.}~\bibnamefont {Xu}}, \bibinfo {author}
  {\bibfnamefont {Y.}~\bibnamefont {Song}}, \bibinfo {author} {\bibfnamefont
  {Y.}~\bibnamefont {Huang}}, \bibinfo {author} {\bibfnamefont
  {D.}~\bibnamefont {Shen}}, \bibinfo {author} {\bibfnamefont {N.}~\bibnamefont
  {Ma}}, \bibinfo {author} {\bibfnamefont {A.}~\bibnamefont {Li}}, \bibinfo
  {author} {\bibfnamefont {S.}~\bibnamefont {Chi}}, \bibinfo {author}
  {\bibfnamefont {M.}~\bibnamefont {Frontzek}}, \bibinfo {author}
  {\bibfnamefont {H.}~\bibnamefont {Cao}}, \bibinfo {author} {\bibfnamefont
  {Q.}~\bibnamefont {Huang}}, \bibinfo {author} {\bibfnamefont
  {W.}~\bibnamefont {Wang}}, \bibinfo {author} {\bibfnamefont {Y.}~\bibnamefont
  {Xie}}, \bibinfo {author} {\bibfnamefont {R.}~\bibnamefont {Zhang}}, \bibinfo
  {author} {\bibfnamefont {Y.}~\bibnamefont {Rong}}, \bibinfo {author}
  {\bibfnamefont {W.~A.}\ \bibnamefont {Shelton}}, \bibinfo {author}
  {\bibfnamefont {D.~P.}\ \bibnamefont {Young}}, \bibinfo {author}
  {\bibfnamefont {J.~F.}\ \bibnamefont {DiTusa}}, \ and\ \bibinfo {author}
  {\bibfnamefont {P.}~\bibnamefont {Dai}},\ }\href {\doibase
  10.1103/PhysRevB.100.094446} {\bibfield  {journal} {\bibinfo  {journal}
  {Physical Review B}\ }\textbf {\bibinfo {volume} {100}},\ \bibinfo {pages}
  {094446} (\bibinfo {year} {2019}{\natexlab{b}})}\BibitemShut {NoStop}%
\end{thebibliography}%

\end{document}